\documentclass[twocolumn,a4paper]{aastex62}
\usepackage{natbib}
\usepackage{amsmath}
\usepackage{longtable}
\usepackage{threeparttable}
\usepackage{footnote}

\newcommand{\source}{MAXI~J1535$-$571}

\newcommand{\degree}{$^{\circ}$}

\newcommand{\arcsecond}{$^{\prime\prime}$}
\newcommand{\perbeam}{\,beam$^{-1}$}

\newcommand{\Swift}{{\it Swift}}
\newcommand{\uJy}{$\mu$Jy}
\newcommand{\mJybeam}{mJy\,beam$^{-1}$}

\newcommand{\swift}{\textit{Swift}}
\newcommand{\lrlx}{$L_{\rm R}$/$L_{\rm X}$}
\newcommand{\lEdd}{$L_{\rm Edd}$}

\graphicspath{{./}{figures/}}

\shorttitle{Disk-jet coupling in \source}
\shortauthors{T. D. Russell et al.}

\begin{document}

\title{Disk-jet coupling in the 2017/2018 outburst of the Galactic black hole candidate X-ray binary MAXI~J1535$-$571}

\correspondingauthor{T. D. Russell}
\email{t.d.russell@uva.nl}

\author[0000-0002-7930-2276]{T. D. Russell}
\affil{Anton Pannekoek Institute for Astronomy, University of Amsterdam, Science Park 904, NL-1098 XH Amsterdam, The Netherlands}

\author{A. J. Tetarenko}
\affiliation{East Asian Observatory, 660 N. A'ohoku Place, University Park, Hilo, Hawaii 96720, USA}
\affiliation{Department of Physics, University of Alberta, CCIS 4-181, Edmonton, AB T6G 2E1, Canada}

\author{J. C. A. Miller-Jones}
\affiliation{International Centre for Radio Astronomy Research - Curtin University, GPO Box U1987, Perth, WA 6845, Australia}

\author{G. R. Sivakoff}
\affiliation{Department of Physics, University of Alberta, CCIS 4-181, Edmonton, AB T6G 2E1, Canada}

\author{A. S. Parikh}
\affiliation{Anton Pannekoek Institute for Astronomy, University of Amsterdam, Science Park 904, NL-1098 XH Amsterdam, The Netherlands}

\author{S. Rapisarda}

\affiliation{Shanghai Astronomical Observatory and Key Laboratory for Research Galaxies and Cosmology, Chinese Academy of Sciences, 80 Nandan Road, Shanghai 200030, People's Republic of China}
\affiliation{Anton Pannekoek Institute for Astronomy, University of Amsterdam, Science Park 904, NL-1098 XH Amsterdam, The Netherlands}

\author{R. Wijnands}
\affiliation{Anton Pannekoek Institute for Astronomy, University of Amsterdam, Science Park 904, NL-1098 XH Amsterdam, The Netherlands}

\author{S. Corbel}
\affiliation{AIM, CEA, CNRS, Universit\'{e} Paris Diderot, Sorbonne Paris Cit\'{e}, Universit\'{e} Paris-Saclay, F-91191 Gif-sur-Yvette, France}
\affiliation{Station de Radioastronomie de Nan\c{c}ay, Observatoire de Paris, PSL Research University, CNRS, Univ. Orl\'eans, 18330 Nan\c{c}ay, France}

\author{E. Tremou}
\affiliation{AIM, CEA, CNRS, Universit\'{e} Paris Diderot, Sorbonne Paris Cit\'{e}, Universit\'{e} Paris-Saclay, F-91191 Gif-sur-Yvette, France}

\author{D. Altamirano}
\affiliation{School of Physics and Astronomy, University of Southampton, Highfield SO17 IBJ, England}

\author{M. C. Baglio}
\affiliation{New York University Abu Dhabi, P.O. Box 129188, Abu Dhabi, United Arab Emirates}

\author{C. Ceccobello}
\affiliation{Department of Space, Earth and Environment, Chalmers University of Technology, Onsala Space Observatory, 439 92 Onsala, Sweden}

\author{N. Degenaar}
\affiliation{Anton Pannekoek Institute for Astronomy, University of Amsterdam, Science Park 904, NL-1098 XH Amsterdam, The Netherlands}

\author{J. van den Eijnden}
\affiliation{Anton Pannekoek Institute for Astronomy, University of Amsterdam, Science Park 904, NL-1098 XH Amsterdam, The Netherlands}

\author{R. Fender}
\affiliation{Astrophysics, Department of Physics, Denys Wilkinson Building, Keble Road, Oxford OX1 3RH, UK}

\author{I. Heywood}
\affiliation{Astrophysics, Department of Physics, Denys Wilkinson Building, Keble Road, Oxford OX1 3RH, UK}
\affiliation{Centre for Radio Astronomy Techniques and Technologies, Department of Physics and Electronics, Rhodes University, Grahamstown 6140, South Africa}

\author{H. A. Krimm}
\affiliation{National Science Foundation, 2415 Eisenhower Ave, Alexandria, VA 22314, USA}

\author{M. Lucchini}
\affiliation{Anton Pannekoek Institute for Astronomy, University of Amsterdam, Science Park 904, NL-1098 XH Amsterdam, The Netherlands}

\author{S. Markoff}
\affiliation{Anton Pannekoek Institute for Astronomy, University of Amsterdam, Science Park 904, NL-1098 XH Amsterdam, The Netherlands}

\author{D. M. Russell}
\affiliation{New York University Abu Dhabi, P.O. Box 129188, Abu Dhabi, United Arab Emirates}

\author{R. Soria}
\affiliation{College of Astronomy and Space Sciences, University of the Chinese Academy of Sciences, Beijing 100049, China}
\affiliation{Sydney Institute for Astronomy, School of Physics A28, The University of Sydney, Sydney, NSW 2006, Australia}

\author{P.~A. Woudt}
\affiliation{Inter-University Institute for Data-Intensive Astronomy, Department of Astronomy, University of Cape Town, Private Bag X3, Rondebosch 7701, South Africa
}

\begin{abstract}
\source{} is a Galactic black hole candidate X-ray binary that was discovered going into outburst in 2017 September. In this paper, we present comprehensive radio monitoring of this system using the Australia Telescope Compact Array (ATCA), as well as the MeerKAT radio observatory, showing the evolution of the radio jet during its outburst. Our radio observations show the early rise and subsequent quenching of the compact jet as the outburst brightened and then evolved towards the soft state. We constrain the compact jet quenching factor to be more than 3.5 orders of magnitude. We also detected and tracked (for 303\,days) a discrete, relativistically-moving jet knot that was launched from the system. From the motion of the apparently superluminal knot, we constrain the jet inclination (at the time of ejection) and speed to $\leq 45^{\circ}$ and $\geq0.69$c, respectively. Extrapolating its motion back in time, our results suggest that the jet knot was ejected close in time to the transition from the hard intermediate state to soft intermediate state. The launching event also occurred contemporaneously with a short increase in X-ray count rate, a rapid drop in the strength of the X-ray variability, and a change in the type-C quasi-periodic oscillation (QPO) frequency that occurs $>$2.5\,days before the first appearance of a possible type-B QPO.

\end{abstract}


\section{Introduction}
Accreting stellar-mass black hole (BH) X-ray binaries (XRBs) launch powerful jets that are observable from radio to infrared (IR) wavelengths (and possibly even up to the X-ray and even $\gamma$-ray band). These jets are capable of carrying away a significant fraction of the accretion power and depositing large amounts of energy into their surroundings \citep[e.g.,][]{2005Natur.436..819G,2007MNRAS.376.1341R,2018MNRAS.475..448T}, that may alter star formation, galaxy evolution, and even the distribution of matter in the Universe \citep[e.g.,][]{1998A&A...331L...1S,2001PhR...349..125B,2011A&A...528A.149M,2012ARA&A..50..455F}. While jet production appears to be fundamentally linked to the process of accretion, the exact nature of the coupling remains poorly understood \citep[e.g.,][]{2004MNRAS.355.1105F}, and how jets are launched, accelerated, and collimated by the accretion inflow is not yet clear.

BH XRBs occasionally go through episodic phases of enhanced accretion (called outbursts) where they brighten significantly as the accretion flow \citep[e.g.,][]{2010LNP...794...53B} and the jets change dramatically \citep[e.g.,][]{2004MNRAS.355.1105F,2004ApJ...617.1272C,2006csxs.book..381F,2014SSRv..183..323F}. These systems evolve through their full outburst cycles on timescales of weeks, months and sometimes years, allowing the full evolution of their accretion and jet duty cycles to be observed in detail. This is one of the reasons that XRBs are excellent laboratories to study BH accretion and jet phenomena. 
 
During the initial rising phase of a typical outburst, BH XRBs are in a hard X-ray spectral state \citep[see, e.g.,][for a review on the accretion states]{2010LNP...794...53B}. This state is characterised by a hard power-law component in the X-ray spectrum \citep[e.g.,][]{1995ApJ...452..710N} and flat or slightly inverted radio spectrum ($\alpha \gtrsim 0$, where the radio flux, $S_{\nu}$, is proportional to the frequency, $\nu$, such that $S_\nu \propto \nu^{\alpha}$; e.g.,\ \citealt{2001MNRAS.322...31F}) from a persistent, partially self-absorbed compact jet \citep[e.g.,][]{2000ApJ...543..373D,2000A&A...359..251C,2001MNRAS.327.1273S}. This flat spectrum extends up to a frequency above which the jet is no longer self-absorbed, and the jet spectrum breaks. At this spectral break frequency (typically at IR frequencies during the beginning of the outburst; \citealt{2013MNRAS.429..815R}), the optically-thick synchrotron spectrum transitions to a steep optically-thin spectrum ($\alpha \approx -0.7$; \citealt{2002ApJ...573L..35C,2013MNRAS.429..815R}). The frequency of this break is related to the distance between the BH and the location where non-thermal particles are first accelerated in the jet \citep[e.g.,][]{2001A&A...372L..25M,2005ApJ...635.1203M,2017SSRv..207....5R,2018MNRAS.473.4417C}, where higher frequencies lie closer to the central compact object.

As the outburst progresses, the accretion rate increases and the X-ray and radio emission continue to brighten. The X-ray spectrum softens as it becomes increasingly dominated by softer (multi-temperature) blackbody emission arising from a geometrically-thin disk. During this softening, the X-ray spectral and variability properties change as the system transits through the hard (HIMS) and soft (SIMS) intermediate states as it moves towards the soft state. 

The transition from the HIMS to the SIMS is typically marked by a rapid decrease in the fractional rms variability of the X-ray emission and the transition between two types of quasi-periodic oscillations (QPOs), type-C and type-B QPOs, respectively (e.g., \citealt{1999ApJ...526L..33W,2001ApJS..132..377H,2002ApJ...580.1030R}, see \citealt{2005ApJ...629..403C} and \citealt{2010LNP...794...53B} for reviews). At some point during this progression, the jet properties change significantly. The steady, compact jet switches off (being quenched by at least 2.5 orders of magnitude in the radio band; \citealt{2011ApJ...739L..19R}), as the jet spectral break evolves to lower frequencies (through the radio band; e.g., \citealt{2013MNRAS.431L.107C,2013ApJ...768L..35R,2014MNRAS.439.1390R}). The transient jet is launched during this phase. The radio emission from this jet is characterised by bright radio flares (e.g.,\ \citealt{2017MNRAS.469.3141T}) that exhibit a steep radio spectrum (e.g.,\ \citealt{2001MNRAS.322...31F}), thought to originate from ejected (optically-thin) synchrotron emitting plasma that collides either with the pre-existing and slower-moving jet, giving rise to internal shocks \citep[e.g.,][]{2010MNRAS.401..394J}, or with the surrounding environment. However, the sequence of the changes in the properties of the accretion flow and jet is currently poorly understood.

The transient jet is composed of discrete, bright, relativistically-moving knots/ejecta that move outwards, away from the compact object \citep[e.g.,][]{2004ApJ...617.1272C,2004MNRAS.355.1105F}. To date, these discrete ejecta have been directly resolved in only a handful of BH LMXB systems \citep[e.g.,][]{1994Natur.371...46M,1995Natur.375..464H,1995Natur.374..141T,1999MNRAS.304..865F,2001ApJ...553..766M,2010MNRAS.409L..64Y,2012MNRAS.419L..49M,2019Natur.569..374M}. The mechanism responsible for the launching of these discrete jet knots is not well understood. Attempts to link changes in the X-ray properties to the timing of the radio flares were not able to identify a clear signature \citep[e.g.,][]{2009MNRAS.396.1370F, 2017MNRAS.469.3141T}. However, there is an expected, but unknown, delay between the launching time and subsequent radio flaring (due to travel time and optical depth effects, as well as the cadence of radio observations typically not detecting the immediate onset of the flare). This delay can be accounted for by tracking the motion of the knot away from the BH and extrapolating back in time to determine the true launching time \citep[e.g.,][]{2009MNRAS.396.1370F,2012MNRAS.419L..49M}. 

The source then transitions to the soft X-ray state, where the X-ray emission is dominated by a multi-temperature blackbody component, with a weak, steep powerlaw component. In the soft state, radio emission from the compact jet is not detected, although some radio emission may be detected from the transient jets as they move downstream and interact with the surrounding environment (e.g., \citealt{2002Sci...298..196C,2004ApJ...617.1272C,2017MNRAS.468.2788R}).

As the accretion rate decreases, the X-ray luminosity decreases and the source begins to spectrally harden. During the reverse transition through the intermediate states and back to the hard state, the compact jet gradually re-establishes itself. The jet is first detected at radio wavelengths, then in the millimetre and infrared bands \citep[e.g.,][]{2012MNRAS.421..468M,2013ApJ...779...95K}, as the jet spectral break shifts to higher frequencies \citep{2013ApJ...768L..35R,2014MNRAS.439.1390R}. Jet flaring is not observed during this re-ignition\footnote{Except in Cyg~X-3, where radio flares during the soft to hard transition are thought to arise when the re-ignited jet has to burrow its way through the channel that has been filled in by the winds from the Wolf-Rayet companion \citep{2010MNRAS.406..307K}.}. The source then typically fades towards quiescence, as the outburst ends.

\subsection{\source{}}

\source{} was first discovered going into outburst when it triggered the Monitor of All-sky X-ray Image (MAXI) Gas Slit Camera (GSC) nova alert system \citep{2010ASPC..434..127N} and the Neil Gehrels Swift Observatory (\swift) Burst Alert Telescope (BAT) hard X-ray transient monitor \citep{2013ApJS..209...14K} on 2017 September 02 (MJD 57998; \citealt{2017ATel10699....1N,2017GCN.21788....1M}). Subsequent X-ray and optical follow-up observations localised the position of the source \citep{2017ATel10700....1K, 2017ATel10702....1S, 2017ATel10704....1S}. \source{} was identified as a BH candidate due to its X-ray \citep{2017ATel10708....1N} and radio \citep{2017ATel10711....1R} properties. Further follow up observations at other wavelengths detected the counterpart in the IR (\citealt{2017ATel10716....1D}) and millimetre (mm; \citealt{2017ATel10745....1T}) bands. From HI absorption, the source distance was estimated to be 4.1$^{+0.6}_{-0.5}$\,kpc \citep{2019arXiv190508497C}.

After the end of its outburst, \source{} did not decay towards quiescence. Instead it exhibited multiple ($>$5), short (and progressively less luminous) X-ray re-brightenings \citep[][]{2018ATel11652....1P,2018ATel11682....1N,2018ATel11884....1L,2019ApJ...878L..28P}. During a few of these re-brightenings, the source transitioned between the hard and soft states, and radio emission was detected (see \citealt{2019ApJ...878L..28P} for discussions of the radio and X-ray observations taken during the re-brightenings).

In this paper, we present comprehensive radio monitoring of \source{} during its outburst, showing the evolution of the jet. In particular, we discuss the quenching of the compact jet, as well as the launching of the transient jet over the hard to soft state transition. We track the motion of a discrete jet knot over 303\,days, which allows us to place constraints on the time of its launching and the properties of the accretion flow at the time of the ejection. We also use the observed properties of the detected jet knot to place constraints on the inclination, speed and opening angle of the jet at the time of the ejection.

While this work only discusses the radio and X-ray behaviour during the major outburst (and not the re-brightenings), we present detections of the jet knot up to $\sim$5\,months after the end of the major outburst.

\section{Observations}
\label{sec:observations}

\subsection{ATCA radio observations}
\label{sec:ATCA}

We monitored the radio counterpart of \source{} with the Australia Telescope Compact Array (ATCA) during its 2017/2018 outburst (under project codes C2604 and C3057). Throughout the major outburst, we carried out observations on 37 epochs between MJD~58001 (2017 September 05) and MJD~58249 (2018 May 11). The ATCA observations were taken every 1--10\,days during the initial hard state rise and transition to the soft state (MJDs~58001 to 58060), every 1--4\,weeks throughout the soft state (MJDs~58060 to 58221), and every 1--10\,days during the decay phase (MJDs~58031 to 58249). 

All observations were taken at central frequencies of either 5.5 and 9.0\,GHz, 17.0 and 19.0 GHz, or at all four frequencies. Each frequency pair (either 5.5 and 9.0, or 17.0 and 19.0\,GHz) was recorded simultaneously with a bandwidth of 2\,GHz at each frequency. Primary flux calibration was done using either PKS~1934$-$638 or PKS~0823$-$500, depending on whether the preferred source, PKS~1934$-$638 was visible at the time of the observation. PKS 1520$-$58 was used for secondary phase calibration for all observations except those taken on MJD\,58001 and MJD\,58008 where 1511$-$55 was used. The data were edited for instrumental issues and radio frequency interference (RFI) before being calibrated following standard routines\footnote{e.g., https://casaguides.nrao.edu/index.php/ATCA\_Tutorials} in the Common Astronomy Software Application (\textsc{CASA} version 5.1.0; \citealt{2007ASPC..376..127M}). Calibrated data were then imaged using \texttt{CLEAN} within \textsc{CASA}. The 5.5\,GHz data were imaged with a robust parameter of 0 to minimise effects due to extended emission from a nearby ($\sim$180$^{\prime\prime}$ away) source. All other frequencies were imaged with a robust parameter of 2 (natural weighting) to maximise the image sensitivity. Where possible (when \source{} was detected above $\gtrsim$10\,mJy), the data were self-calibrated (phase and amplitude) down to a solution interval of 10\,seconds.  

We determined the radio flux density ($S_{\rm R}$) by fitting a point source in the image plane. All flux densities are reported in Table~\ref{tab:ATCA_data} and shown in Figure~\ref{fig:lc}. The radio luminosity ($L_{\rm R}$) was calculated by $L_{\rm R}$=$4\pi S_{\rm R} \nu D^2$, where $D$ is the source distance. 

We measured the position of the jet knot, the core position of \source (hereafter, only referred to as \source), and other objects in the field by
fitting point sources in the uv-plane of the 9\,GHz\footnote{The 9\,GHz observations provide the best balance between sensitivity, resolution, and phase stability. Position fitting was carried out before self-calibration.} observations using \textsc{UVMULTIFIT} \citep{2014A&A...563A.136M}. For each epoch, the positions of the target and jet knot were corrected using the positional offsets determined from a bright source in the field. All applied positional shifts were $<$0.5\arcsecond{} in Right Ascension and $<$0.8\arcsecond{} in Declination, with both being typically $<$0.2\arcsecond. Measured flux densities and positions (as well as the telescope configuration) are presented in Table~\ref{tab:ATCA_data_S2} and Table~\ref{tab:S2_positions}, respectively.

\subsubsection{Intra-observation variability}
\label{sec:core_variability}

\textsc{UVMULTIFIT} was also used to search for source variability within each epoch, where we fit for a point source in the uv-plane for different time intervals within each observation. Many of our radio observations were short in duration (generally only 15--30\,mins long), meaning that for many observational epochs the short-time variability could not be well explored. Therefore, for most ATCA observations during the outburst we only observed small changes in the source flux density (by $\lesssim$3\%) over the radio observation. 

However, we detected source variability on three epochs, on MJDs~58013, 58017 and 58019, where the source varied significantly when compared to other sources in the field. This variability is discussed in Section~\ref{sec:results} and shown in Figures~\ref{fig:variability} and \ref{fig:variability_58019}. 

\subsection{MeerKAT radio observation}
\label{sec:MeerKAT}

The field surrounding MAXI J1535-571 was observed with the MeerKAT radio observatory \citep{2018Camilo,2016jonas} for 2.1\,hours on MJD~58222 (2018 April 14), as part of the ThunderKAT Large Survey Project \citep{2017fender}. The observation was recorded at a central frequency of 1.28\,GHz with a bandwidth of 856\,MHz split into 4096~channels, and an 8~second integration time. PKS~0408$-$658 was used for bandpass and flux calibration, while PKS~1421$-$490 was used for phase calibration.

The data were flagged using AOFlagger\footnote{https://sourceforge.net/projects/aoflagger/} (version 2.9; \citealt{2010offringa}) and calibrated following standard procedures within \textsc{CASA} (version 5.1.0, \citealt{2007ASPC..376..127M}). To reduce data volume the raw data was binned (8 channels per bin), resulting in 512 channels with a channel-width of 1.67~MHz each. Imaging was then carried out with the new wide-band, wide-field imager, DDFacet \citep{2018tasse}. DDFacet is based on a co-planar faceting scheme and takes into account generic direction-dependent effects that dominate wide fields (such as the $\sim$1$^{\circ}$ field-of-view of MeerKAT). A Briggs robust parameter of 0 was used during imaging, and deconvolution was done over four frequency blocks using the \textsc{SSDclean} deconvolution algorithm. DDFacet is accompanied by the calibration software killMS\footnote{https://github.com/saopicc/killMS}, which was used to self-calibrate the data in order to correct for artifacts from bright sources. The image quality was also optimized using the \textsc{CohJones} (Complex Half-Jacobian Optimization for N-directional EStimation; \citealt{2015tasse}) algorithm, which solves for scalar Jones matrices in a user-defined number of directions and includes corrections for direction-dependent effects.

The position of the jet knot was measured before self-calibration using \textsc{imfit} within \textsc{CASA}. As described in Section~\ref{sec:ATCA}, we corrected the position relative to the location of the same bright nearby source used for the 9\,GHz ATCA data. The measured flux density and position are given in Table~\ref{tab:ATCA_data_S2} and Table~\ref{tab:S2_positions}, respectively.

\subsection{X-ray observations}
\label{sec:Swift}

\source{} was well monitored in the X-ray band throughout its major outburst. The \swift-X-ray telescope (XRT) monitored \source{} (target ID: 00010264) during the outburst rise and decay, MAXI observed \source{} intensively throughout the entire outburst, while HXMT \citep{2018ApJ...866..122H} and AstroSAT \citep{2019MNRAS.487..928S,2019arXiv190610595B}  monitored the source densely for periods during the rise of the outburst. 

For our comparison between the hard state radio ($L_{\rm R}$) and X-ray ($L_{\rm X}$) luminosities (Section~\ref{sec:lrlx}), we analysed ATCA observations of \source{} and \swift-XRT observations of \source, respectively. This X-ray analysis is reported in full detail by \citet{2019ApJ...878L..28P}, but we briefly summarise the analysis below. 

All \swift-XRT data were downloaded from the \textsc{HEASARC} archive and processed using \texttt{xrtpipeline}. Pile-up corrected \swift-XRT Windowed-timing mode observations were extracted in the 0.7--10\,keV range and then fit using the X-ray spectral fitting package (\textsc{XSPEC}, version 12.9.1; \citealt{1996ASPC..101...17A}). The equivalent hydrogen column density ($N_{\rm H}$) was modelled using \texttt{WILM} abundances \citep{2000ApJ...542..914W} with \texttt{tbabs} and \texttt{VERN} cross-sections \citep{1996ApJ...465..487V}. $N_{\rm H}$ was left as a free parameter, where the value used for the X-ray luminosities was the average, providing $N_{\rm H}$=3.54$\pm$0.03$\times$10$^{22}$\,cm$^{-2}$. To determine the X-ray flux of \source{} at the time of the radio observations, the hard state data were modelled (and well fit) with a simple absorbed powerlaw model (\texttt{tbabs}$\times$\texttt{powerlaw}); the 1--10\,keV X-ray de-absorbed flux ($S_{\rm X}$) was then calculated using the \textsc{XSPEC} convolution model \texttt{cflux} before being converted to a luminosity as $L_{\rm X}$=$4\pi S_{\rm X} D^2$.

\section{Results}
\label{sec:results}

X-ray and radio observations (Figure~\ref{fig:lc}) show \source{} brighten and fade over its 2017/2018 outburst. During this outburst, the source evolved through the X-ray spectral states, producing significant changes in the observed X-ray and radio properties. 

\subsection{X-ray spectral state evolution from the literature}
\label{sec:states}

For the state transitions and general behaviour, we adopt X-ray spectral results reported by \citet{2018MNRAS.480.4443T} based on the \swift-XRT monitoring. However, at times when \swift{} did not observe the source, we also refer to the behaviour and state transitions reported from MAXI monitoring \citep{2018PASJ...70...95N}. The evolution is supported by timing results from HXMT monitoring \citep{2018ApJ...866..122H}. Table~\ref{tab:states} summarises the X-ray spectral state evolution of \source{} from these studies. 

\begin{table}
\caption{Tabulated X-ray spectral state evolution of \source{} during its 2017/2018 major outburst. }
\centering
\label{tab:states}
\begin{tabular}{ll}
\hline
MJD & X-ray state transition\\
\hline

58004.49$^a$--58007.27$^a$  & Hard state $\rightarrow$ HIMS \\
58014.18$^a$--58015.37$^a$ & HIMS $\rightarrow$ SIMS \\
58044$^b$ & IMS $\rightarrow$ Hard state \\
58054$^b$ & Hard state $\rightarrow$ IMS \\
58060$^b$ & IMS $\rightarrow$ Soft state \\
58233$^a$ & Soft state $\rightarrow$ IMS \\
58237$^a$ & IMS $\rightarrow$ Hard state \\

\hline
\multicolumn{2}{l}{  
\begin{minipage}{0.9\columnwidth} \footnotesize
The X-ray spectral evolution is from \swift-XRT monitoring \citep{2018MNRAS.480.4443T} and MAXI \citep{2018PASJ...70...95N}. HIMS and SIMS denote the hard and soft intermediate state, while IMS is the intermediate state from the MAXI monitoring, which does not clearly distinguish between the HIMS and SIMS.\\
$^{a}$ From \citep{2018MNRAS.480.4443T}.\\
$^{b}$ From \citep{2018PASJ...70...95N}.\\

\end{minipage}
}
\end{tabular}

\end{table}

\subsection{Radio results}

Throughout its outburst, we detected radio emission from \source{} that was consistent with either a steady, compact jet or a flaring, transient jet (identified by the flat-to-inverted, or steep radio spectrum, respectively). We also monitored a downstream radio knot from the transient jet that was spatially-resolved from \source{} and moving away. We refer to this discrete knot as S2 hereafter. For clarity, in the results section we present the results from each of these two components separately. Section~\ref{sec:radio_core} describes the radio emission that was spatially coincident with \source, regardless of whether it originated from the compact or transient jet. Section~\ref{sec:ejecta} presents the emission from S2.

\subsection{Radio emission from the location of \source}
\label{sec:radio_core}

Following the initial MAXI and \Swift/BAT detections on MJD~57998 \citep{2017ATel10699....1N,2017GCN.21788....1M}, we observed \source{} with ATCA on MJD\,58001, as it brightened during the hard X-ray state. These ATCA radio observations significantly detected the radio counterpart of \source{} at a Right Ascension (R.A.) and Declination (Dec) of:\\
\\
R.A. (J2000) = 15$^{\rm h}$35$^{\rm m}$19.71$^{\rm s}$ $\pm$ 0.08\\
Dec (J2000) = -57$^{\circ}$13$^{\prime}$47.58$^{\prime \prime}$ $\pm$ 0.06,\\
\\
where the errors presented are the estimated systematics (as a distance from the phase calibrator) added in quadrature with the statistical errors on the fit (which are larger than the theoretical error of beam centroiding, beam/2$\times$SNR). 

\subsubsection{Radio lightcurves and spectra}
\label{sec:lightcurves}

\begin{figure*}[!ht]
\centering
\includegraphics[width=0.94\textwidth]{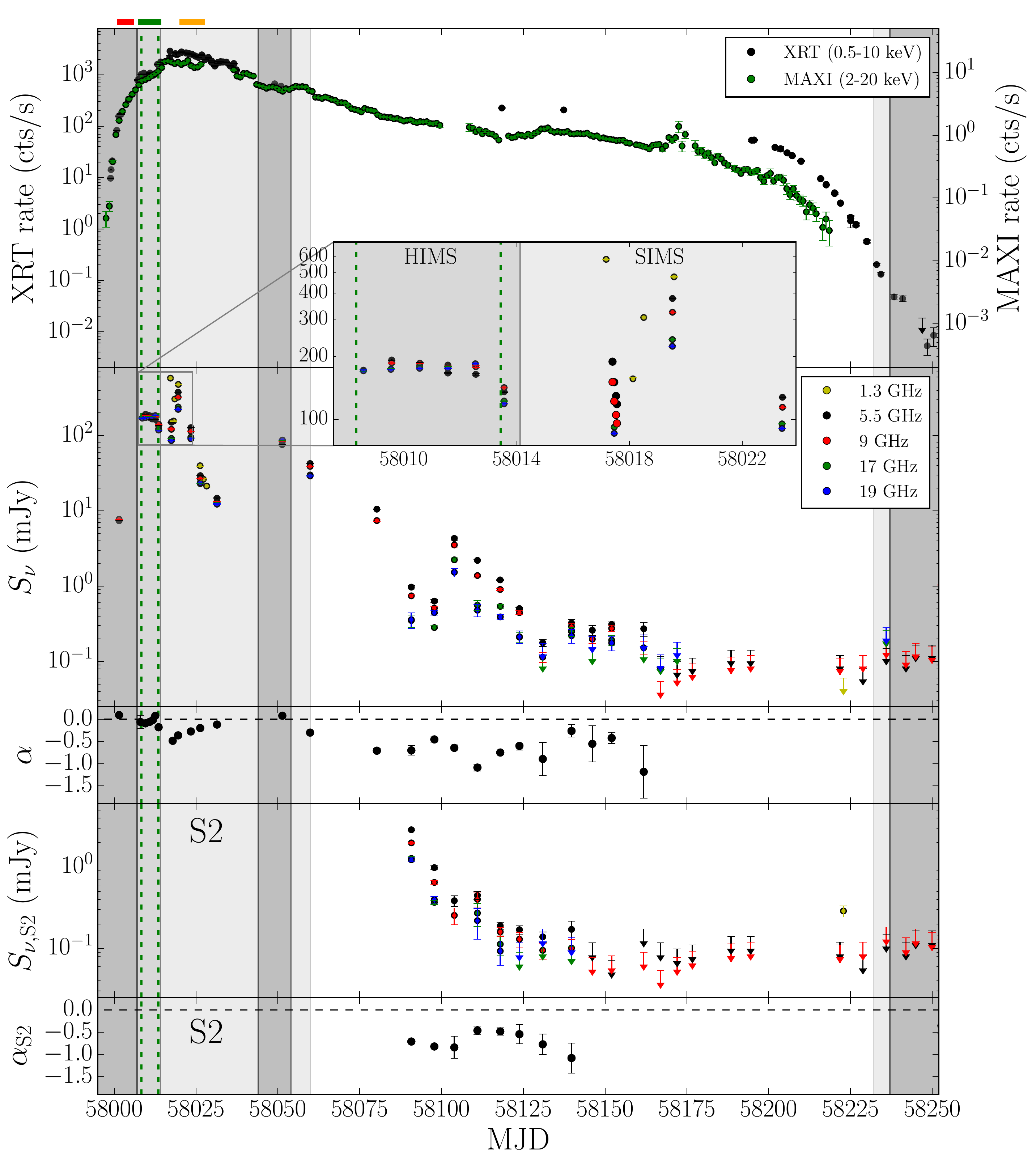}
\caption{X-ray and radio lightcurves of \source{} during its 2017/2018 major outburst. {\it First panel:} Swift/XRT (left axis, black points) and MAXI (right axis, green points) lightcurve of \source{} throughout the outburst. {\it Second panel:} Multi-frequency radio monitoring of \source. Also shown is a zoomed insert of the time range MJD\,58008--MJD\,58023, highlighting the period of jet quenching and flaring. We include 1.34\,GHz ASKAP data from \citep{2019arXiv190508497C} and the short time variability of our observation on MJD~580017.5 showing the fading radio emission from the first radio flare, see also Figure~\ref{fig:variability}. {\it Third panel:} Radio spectral index of \source. {\it Fourth panel:} Flux density measurements of the discrete jet knot, S2. {\it Fifth panel:} Radio spectral index of S2. The dark shaded regions indicate when the source was in a hard state, the lighter shaded regions indicate the progressive intermediate states, where the darker is the HIMS and the lighter is the SIMS. The un-shaded regions indicate the soft states (see Section~\ref{sec:states} and Table~\ref{tab:states} for the timing of the spectral state transitions). We show the extent of the modelled ejection dates at the top of the figure. Red represents the bulk motion model, orange is the decelerated motion model, and green shows the bulk plus decelerated motion model. The green vertical dashed lines show the most plausible ejection dates across all panels. All flux densities are provided in Table~\ref{tab:ATCA_data} and Table~\ref{tab:ATCA_data_S2}. Here we show the evolution of the radio emission from \source{} and S2 during the major outburst.}
\label{fig:lc}
\end{figure*}

Following our initial radio detection on MJD~58001, \source{} brightened at both radio and X-ray wavelengths over the next $\sim$week as the X-ray spectrum slowly softened (Figure~\ref{fig:HID}), although the source remained in the HIMS. After this phase of rapid radio brightening, the radio emission then remained relatively steady ($S_{\rm 5.5\,GHz} \approx 170-180$\,mJy) from $\sim$MJD~58008.5 until MJD~58012.5 (during which we observed daily; Figure~\ref{fig:lc}). Over this time, although the radio spectrum evolved marginally, the spectral index generally remained flat to inverted, indicating the continued presence of the compact jet. 

\begin{figure}[ht]
\centering
\includegraphics[width=0.98\columnwidth]{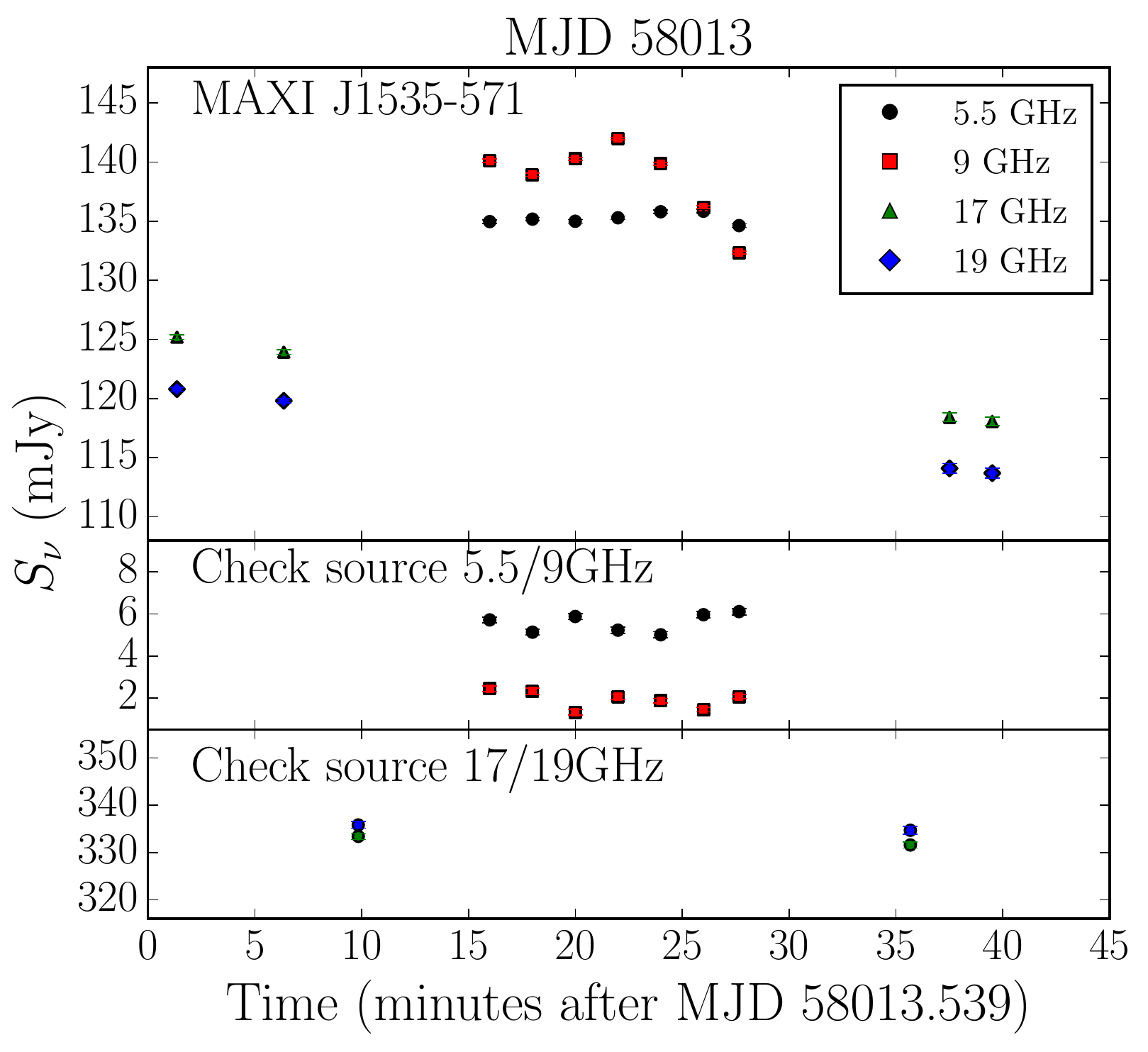}
\includegraphics[width=0.98\columnwidth]{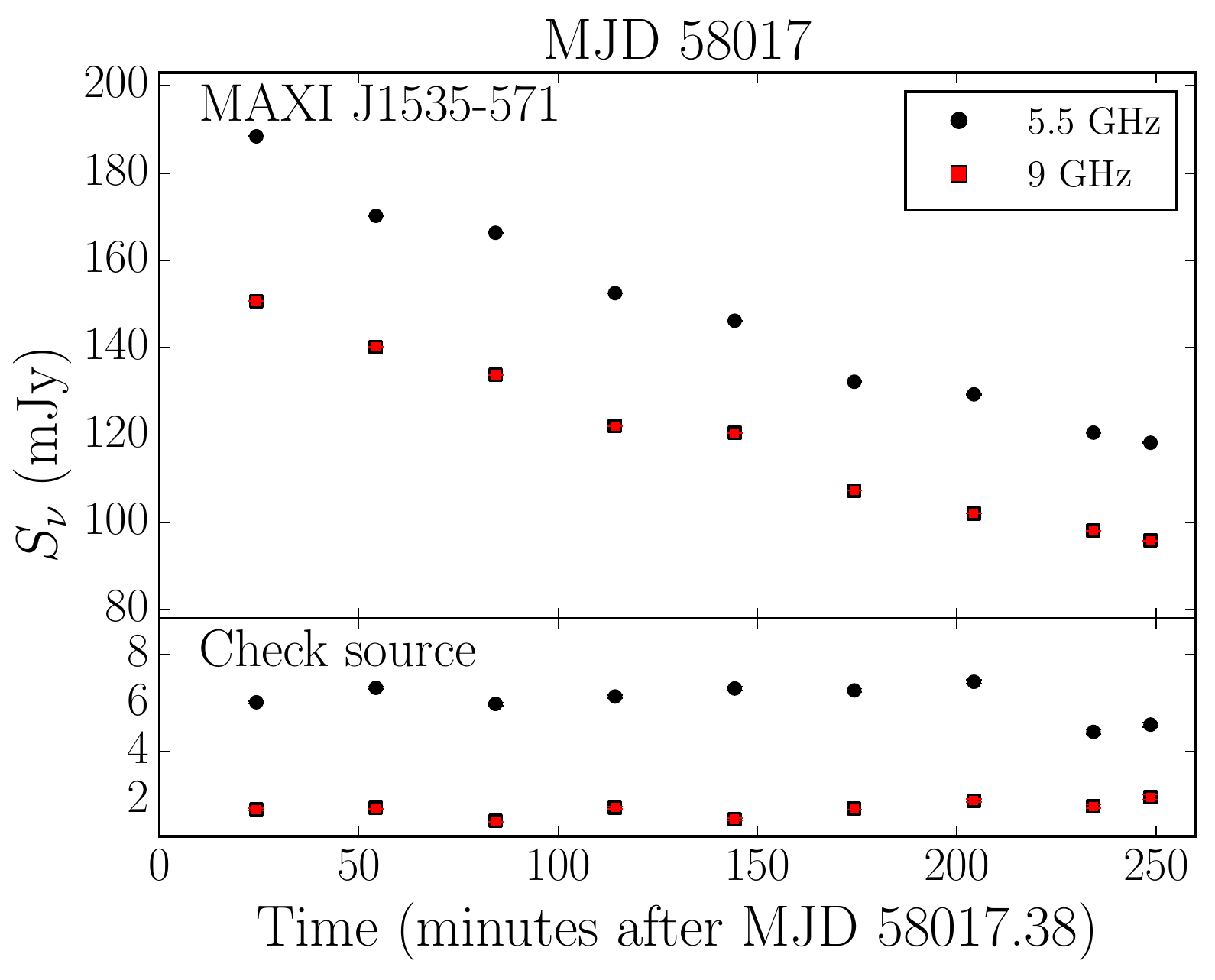}
\caption{Intra-observation radio variability of \source{} during ATCA observations on MJD~58013 (top figure) and MJD~58017 (lower), when the source was in the HIMS and SIMS, respectively. These lightcurves highlight the radio variability of the target. We also show the radio emission from a check source within the field at 5.5 and 9\,GHz, and treat two phase calibrator scans as a target/check source at 17 and 19\,GHz (there was no nearby check source at 17 and 19\,GHz) to show that the variability is intrinsic to \source. The variability on MJD~58013 implies the beginnings of jet quenching as the jet spectrum changes. The fading emission on MJD~58017 indicates the end of the first radio flare.}
\label{fig:variability}
\end{figure}

Our final ATCA observation within the HIMS (starting $\approx$MJD~58013.5) showed that the radio emission as a whole had faded ($S_{\rm 5.5\,GHz} \approx 135$\,mJy) and the radio spectrum had steepened while the X-ray luminosity increased (Figure~\ref{fig:lc}). This epoch is a little more complicated than just steady fading and shows strong evidence that the jet spectral break had moved into, and through, the radio band during this epoch. The evolution of the jet spectral break will be discussed in greater detail in Russell et al. (in preparation), however, we summarise the behaviour during this epoch here. This evidence is highlighted by the radio spectrum initially remaining $\sim$flat(-to inverted) between the 5.5 and 9\,GHz ATCA bands, while the 17 to 19\,GHz emission was fainter than the lower-frequency observing bands (and continuing to fade\footnote{The 17 and 19\,GHz bands were not observed simultaneously with the 5.5 and 9\,GHz bands. Additionally, there was no nearby check source detected at 17 and 19\,GHz, so the intrinsic variability check was done by treating the inner scans of the phase calibrator as a target source, while the other scans were used as the calibrator.}) and exhibited a steep radio spectrum (Figure~\ref{fig:variability}, top). Towards the end of the observation, the 5.5 and 9\,GHz radio spectrum also steepened due to rapid fading at 9\,GHz. This evolution implies that the jet spectral break resided within the radio band during this observation. Such behaviour indicates the beginning of the quenching of the radio jet, which is supported by a sharp drop in the IR and optical brightness between MJD~58012 and MJD~58016 \citep[][]{2018ApJ...867..114B}.

\begin{figure}[ht]
\centering
\includegraphics[width=0.98\columnwidth]{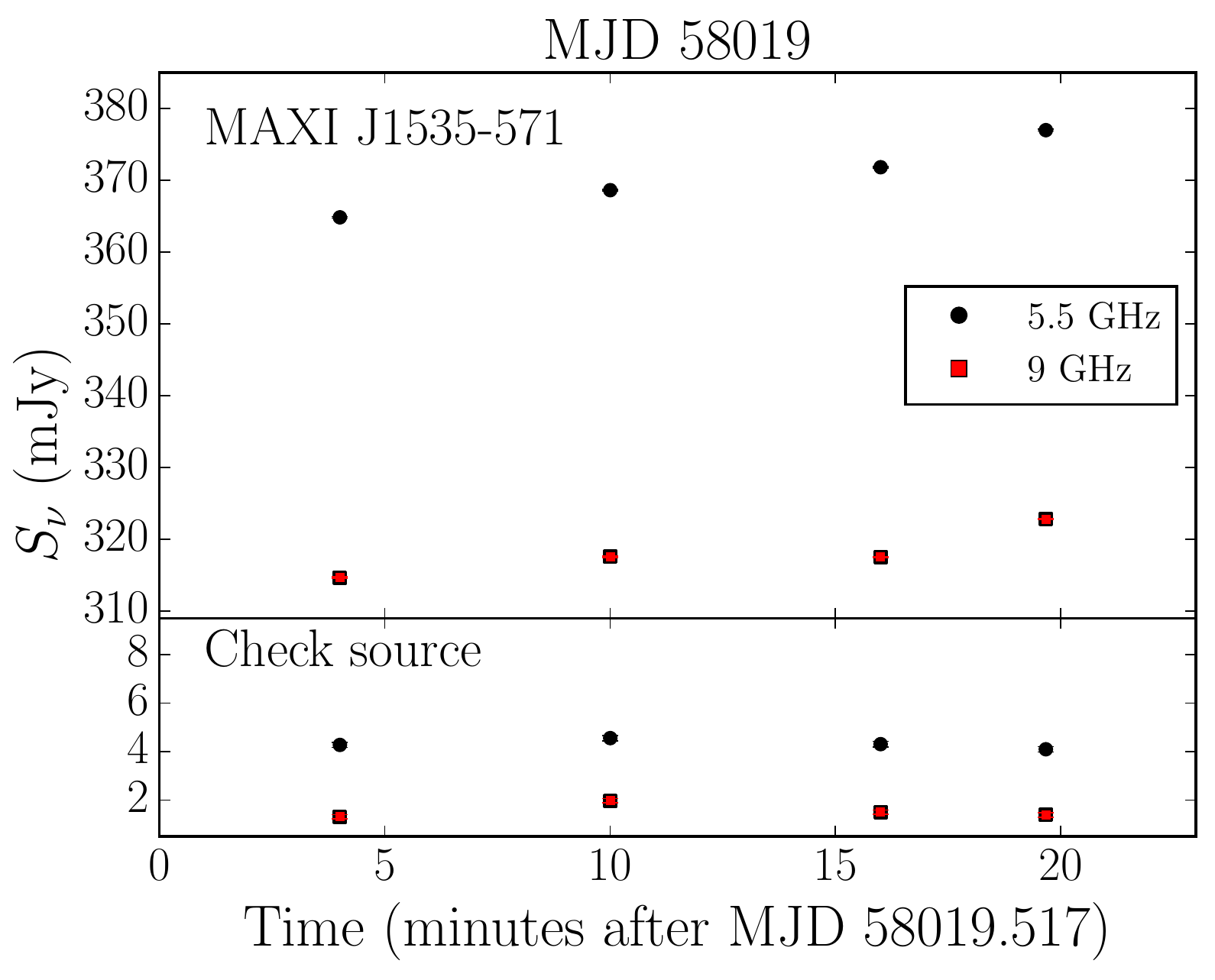}
\caption{Intra-observation radio variability of \source{} (in the SIMS) during ATCA observations on MJD~58019. Here we show the rising radio emission from \source{} during the second radio flare (top panel). We also show the non-variable radio emission from a check source within the field (lower panel). }
\label{fig:variability_58019}
\end{figure}

\begin{figure}[!ht]
\centering
\includegraphics[width=1\columnwidth]{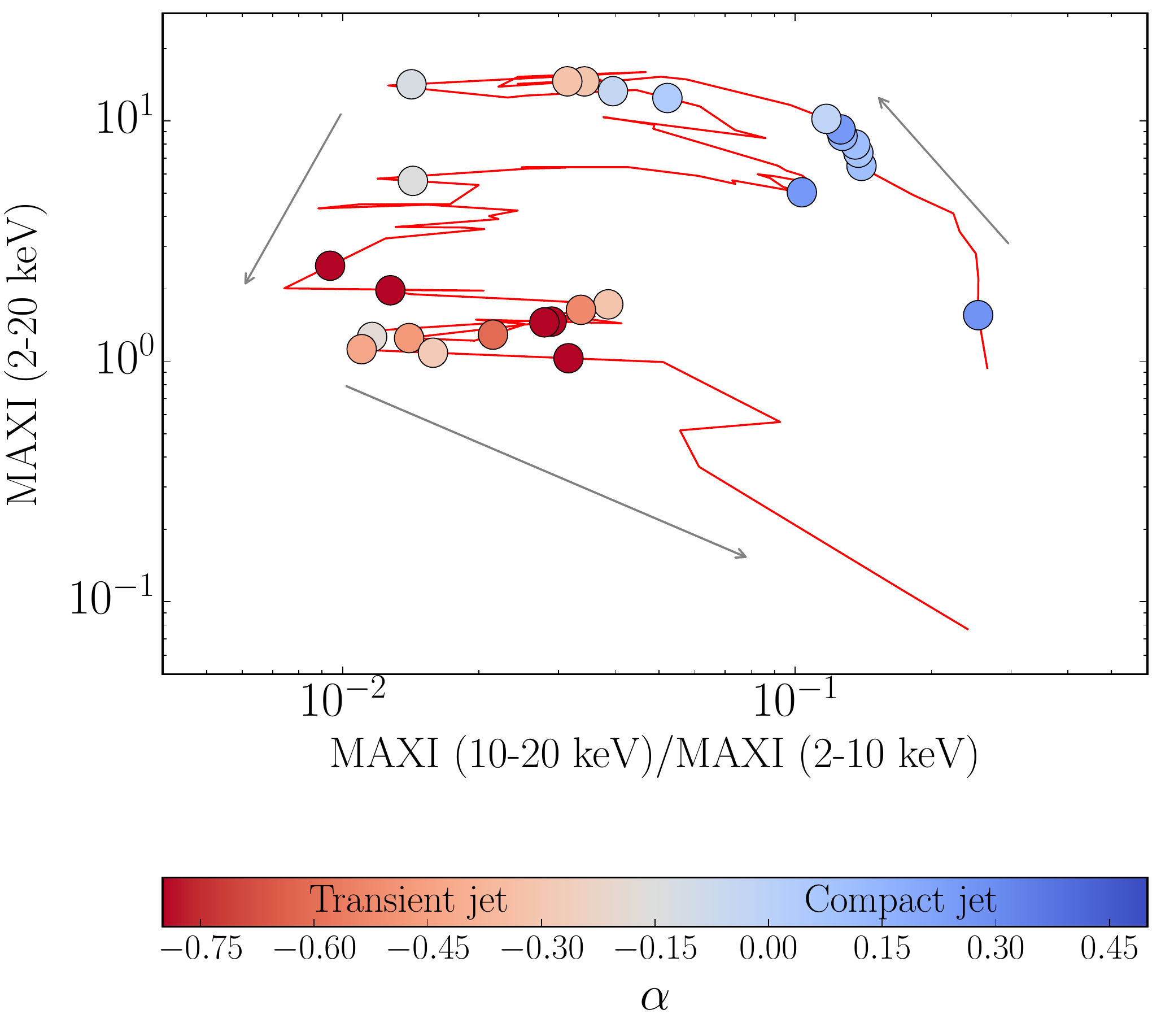}
\caption{Hardness intensity diagram (HID) of \source{} during its major outburst, where the hardness was determined from MAXI data. For clarity, we only show MAXI detections greater than 1.5-$\sigma$. Grey arrows indicate the overall evolution. The timing and spectral index, $\alpha$, of the radio observations are shown by the overlayed circles, where the blue represents a flat-to-inverted radio spectrum (compact jet), and red represents a steep radio spectrum (from the transient jet), as indicated by the colourbar. We see the compact jet evolve to the transient jet as the source underwent rapid X-ray spectral softening. The compact jet recovered during a brief return to the hard state before rapid softening once again (as the radio spectrum became steep).}
\label{fig:HID}
\end{figure}

Our next radio observation ($\approx$MJD~58017.4) occurred within the SIMS. Dividing that radio observation into 2-minute time intervals showed that \source{} was brighter ($S_{\rm 5.5\,GHz} \approx 190$\,mJy) at the beginning of that observation, with the radio emission fading steadily at all frequencies (to $S_{\rm 5.5\,GHz} \approx 118$\,mJy; Figures~ \ref{fig:lc} and \ref{fig:variability}) during the $\approx$3.5-hour radio observation. At all times the radio spectrum was steep ($\alpha \approx -0.5$). Additionally, the Australian Square Kilometre Array Pathfinder (ASKAP) also detected a bright radio peak ($\approx$580\,mJy) in the 1.4\,GHz lightcurve $\sim$MJD~58017.17 \citep{2019arXiv190508497C}, which are included in Figure~\ref{fig:lc}. These results indicate that a bright radio flare occurred between MJD~58013.5 and 58017.4. 

A second bright, steep-spectrum radio flare was observed to be rising during our radio observation on MJD~58019 (Figure~\ref{fig:variability_58019}). This second flare then faded over the next few weeks. However, instead of continuing to fade, \source{} temporarily returned back to a hard state between MJD\,58044 and MJD\,58054. During this period the radio jet re-brightened and the radio spectrum was observed to be flat (Figure~\ref{fig:lc}). 

After this brief return to the hard state, \source{} entered the soft state and faded slowly in the X-ray band over the next $\sim$160\,days. During this steady soft-state decay, we continued to detect radio emission (with generally decreasing flux density and a steep radio spectrum) until MJD~58161, more than 100 days after the hard-to-soft state transition. Throughout this evolution, we observed some minor radio flaring and changes to the radio spectrum (although $\alpha$ was always $<$0), indicating that the emission originated from a transient jet. 

During the decay period at the end of the major outburst, we did not detect radio emission from the source. Of particular interest, during and after the soft to hard state transition (MJD~58237) at the end of the outburst, we detected no radio emission from the source down to 3-$\sigma$ upper-limits of $\approx$120\,\uJy\,beam$^{-1}$.

\subsection{Radio emission associated with the resolved jet knot, S2}
\label{sec:ejecta}

\begin{figure}
\centering
\includegraphics[width=\columnwidth]{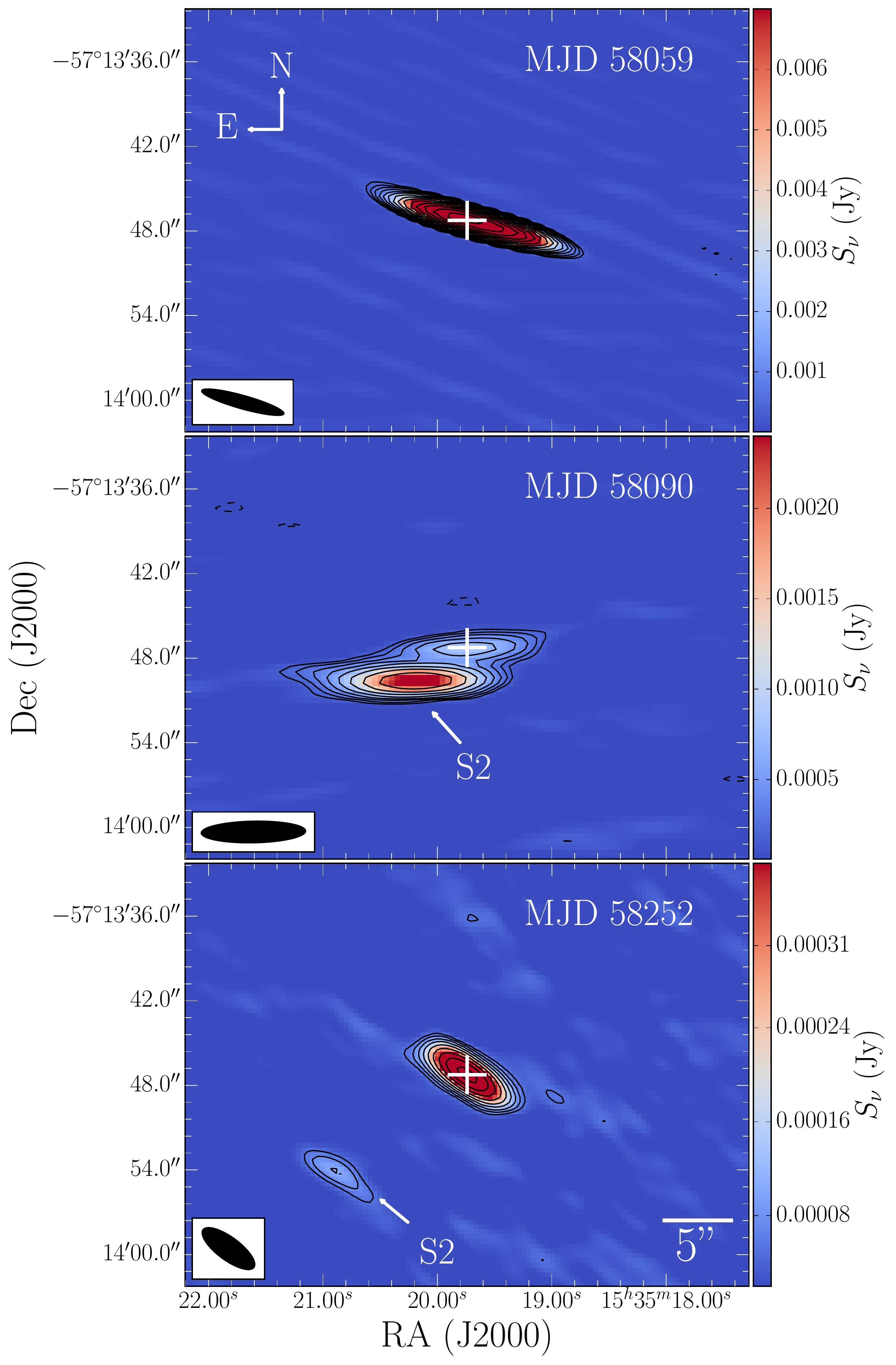}
\caption{A sample of the 5.5\,GHz ATCA monitoring of \source{} and S2, showing the motion of the S2 as it moved away from the source. The \textit{top panel} shows non-detection of S2 on MJD~58059, just after the telescope reconfiguration, the \textit{second panel} shows our initial detection of S2, while the \textit{third panel} shows the detection of S2 at much later times. This sample of images highlights the motion of S2 as it moved away from \source{} (where the source position is marked by the white cross in each image). Contours are $\pm \sqrt{2}^n$ times the rms, where $n$=3,4,5,6... and the rms was 60, 50, and 18\,\uJy\perbeam, respectively. Dashed contours represent negative values. S2 travelled away from \source{} at a position angle of 124.7$\pm$0.5\degree East of North. S2 was detected over 12 observing epochs in total (Table~\ref{tab:ATCA_data_S2}, Figures~\ref{fig:lc} and ~\ref{fig:ejecta_motion}).}
\label{fig:ejectaimage}
\end{figure}

ATCA observations taken on MJD~58090 (2017 December 03) showed a second ($\sim$3\,mJy) radio source (S2) located $\approx$5$^{\prime\prime}$ to the South East of the measured position of \source{} (Figure~\ref{fig:ejectaimage}, middle panel). No counter-jet component was detected.

Imaging and uv-plane analysis for all observations prior to the detection on MJD~58090 did not appear to show any detectable emission significantly offset from the position of \source. In most cases it is not surprising that S2 (or a counter-component) was not spatially-resolved in these earlier observations due to the poor spatial resolution (arising from the compact, H168, telescope configuration providing a spatial resolution of $\sim$160$^{\prime\prime}$ at 5.5\,GHz at those times). However, in early 2017 November, ATCA moved back to its extended 6\,km array configuration, providing spatial resolutions of $\sim$1--5$^{\prime\prime}$. Using the proper motion we measured for S2 (Table~\ref{tab:ejecta_modelling}), it would only have reached angular separations of a few arcseconds $\sim$50--100\,days after the ejection event occurred. Our two ATCA observations taken during 2017 November (on MJDs~58059 and 58080) may have been taken when S2 had travelled sufficiently far from \source, but S2 was not detected in those observations (Figure~\ref{fig:ejectaimage}). On MJD~58059, the lack of detection could be due to S2 still being too nearby to \source{} for it to be spatially resolved, while our observation on MJD~58080 only comprised 2$\times$10\,min on-source scans over a $\sim$40\,min period, providing poor uv-coverage and a high noise level (an rms of $\sim$120\,\mJybeam at 5.5\,GHz and $\sim$170\,\mJybeam at 9\,GHz).

The radio spectrum of S2 remained optically-thin for all detections (Figure~\ref{fig:lc}, fifth panel), and S2 was detected during all ATCA observations over the next $\sim$2\,months, and then sporadically by ATCA over a $\sim$1\,year period, as well as during the single MeerKAT observation on MJD~58222 (Figure~\ref{fig:lc} and Table~\ref{tab:ATCA_data_S2}). We observed S2 moving away from \source{} at a position angle of 124.7$\pm$0.5\degree{} East of North, under the standard assumption of a ballistic (linear) trajectory. 

From its motion and radio spectrum, S2 was consistent with emission from synchrotron emitting plasma, arising from shocks or interactions by a discrete jet knot that was launched from, and subsequently moved away from, \source. At no time was S2 observed to be extended or was another ejected component detected, either travelling in the opposite direction from the target (to the North West of \source) or in the same direction as S2 but at a different separation.

\begin{figure*}[!ht]
\centering
\includegraphics[width=0.9\textwidth]{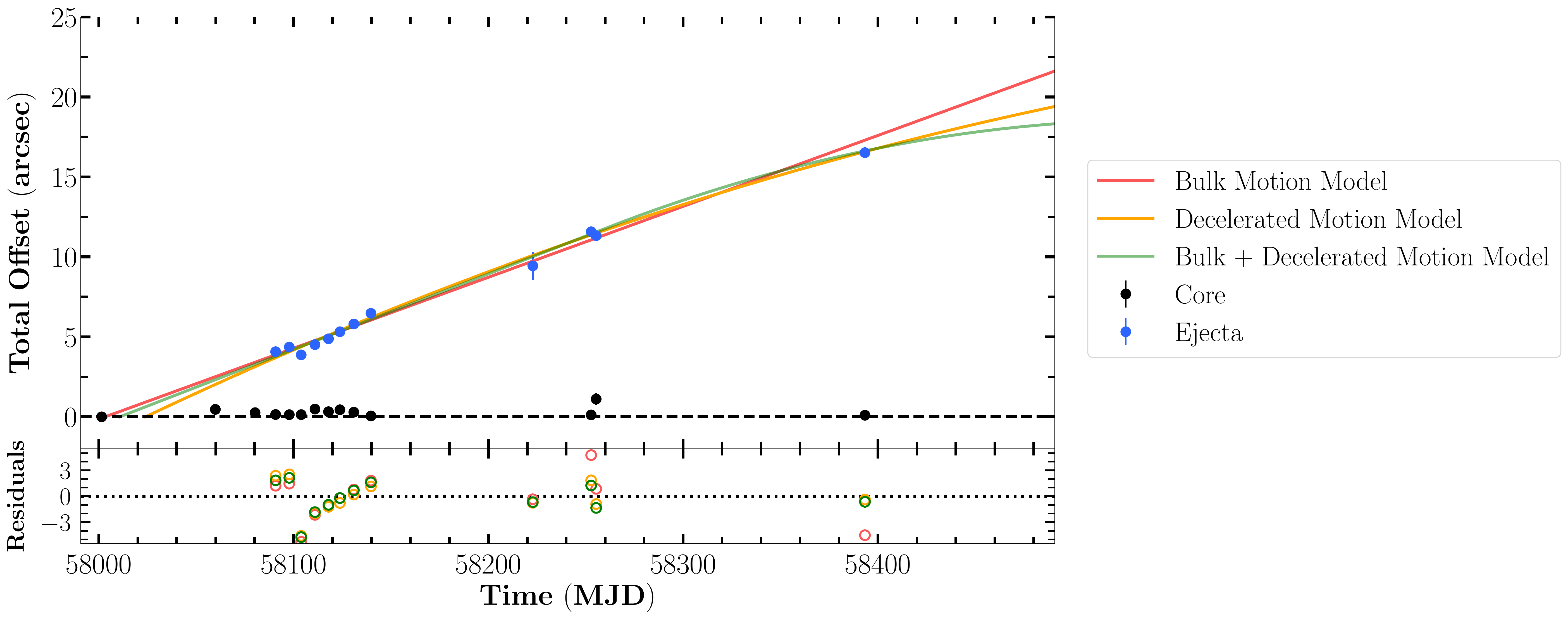}
\caption{The separation (in arcseconds) over time of S2 (blue points) from \source{} (black points). Here, both sources have been normalised to the radio position of \source{} and corrected relative to another bright radio source in the field. We modelled the motion of the S2 with an MCMC algorithm using a constant bulk motion model (red line), a simple decelerating motion model (orange line), and a combination of the two, where S2 initially travelled with a constant velocity, before decelerating at later times. Residuals are shown in the lower panel, which were calculated as the data minus the model divided by the observational uncertainties. Extrapolating the motion of the knot back in time, these models estimate the time of zero separation between the core and knot (time of launching) to be MJD~58003.4$^{+1.6}_{-1.7}$, MJD~58024.1$^{+2.6}_{-3.2}$, or MJD~58010.8$^{+2.65}_{-2.5}$, respectively. Full model results are shown in Table~\ref{tab:ejecta_modelling}.}
\label{fig:ejecta_motion}
\end{figure*}

S2 faded steadily following its initial detection (Figure~\ref{fig:lc}), dropping below our detection threshold $\sim$50\,days after its initial identification (on MJD~58090). Over this time, the radio spectrum remained steep, although it did vary.

During our continued monitoring of \source{} (during the reflares the source displayed after the major outburst ended), two additional brightenings of S2 were detected, allowing its motion away from the black hole to be tracked over a period of 303\,days during our monitoring.

To fit the motion of S2 we use a MCMC algorithm (\textsc{emcee}; \citealt{2013PASP..125..306F}), where the best fit result is taken as the median of the one-dimensional posterior distributions, and the uncertainties are reported as the range between the median and the 15th percentile ($-$ve), and the 85th percentile and the median ($+$ve), corresponding approximately to 1-$\sigma$ errors. Results from the MCMC fitting are reported in Table~\ref{tab:ejecta_modelling} and shown in Figure~\ref{fig:ejecta_motion}. We show the parameter correlations in the supplementary information (Figures~\ref{fig:pc1}, \ref{fig:pc2}, and \ref{fig:pc3}), which show no bi-modal posterior distributions. We note that the small decrease we measure in the separation of S2 from \source{} on MJD~58103 (Figure~\ref{fig:ejecta_motion}) likely arises from poor uv-coverage due to the short observation length resulting in unaccounted for systematics for that epoch.

While the majority of our monitoring appears to show S2 moving away from \source{} linearly in time (i.e., with a constant velocity; Figure~\ref{fig:ejecta_motion}), the final measured position seems to have not travelled as far as expected. This implies S2 decelerated over time, either for all times following its ejection, or just at later times. Therefore, we model the proper motion of S2 in three different ways. In the first case, we assume S2 travelled with a constant bulk motion (constant velocity). For the second case, we include an acceleration component in our model, allowing S2 to decelerate from the moment of launching. In the third case, we combined these two scenarios; where S2 initially travelled with a constant velocity until a time, $t_{\rm decel}$, following which, S2 could decelerate. Such late-time deceleration could occur due to interactions with the ISM, the pre-existing jet, or once the knot had swept up enough mass to slow itself (see Section~\ref{sec:S2_discussion} for further discussion).

For case of constant motion, we describe the motion of S2 away from the source as:
\begin{equation}
\begin{split}
{\rm R.A.}_{\rm offset}&=\mu_{\rm R.A.}\left(t-t_{\rm ej, bulk}\right),\\
{\rm Dec}_{\rm offset}&=\mu_{\rm Dec}\left(t-t_{\rm ej, bulk}\right),
\end{split}
\label{eq:bulkmotion}
\end{equation}
where R.A.$_{\rm offset}$ and Dec$_{\rm offset}$ correspond to the positional offset from the location of \source{} in R.A. and Dec., respectively, $\mu_{\rm R.A.}$ and $\mu_{\rm Dec}$ are the proper motions in mas\,day$^{-1}$, $t$ is the time (in MJD), and $t_{\rm ej, bulk}$ is the time of zero separation between S2 and the source (time of ejection). Fitting all data points without any weighting determines the time of ejection, $t_{\rm ej, bulk}$, to be MJD~58003.4$^{+1.6}_{-1.7}$. Full results are shown in Table~\ref{tab:ejecta_modelling}.

Including constant deceleration at all times, we describe the motion away from the source as:
\begin{equation}
\begin{split}
{\rm R.A.}_{\rm offset}&=\mu_{\rm R.A.}\left(t-t_{\rm ej, decel}\right)-\frac{1}{2} \dot{\mu}_{\rm R.A.}\left(t-t_{\rm ej, decel}\right)^{2}, \\
{\rm Dec}_{\rm offset}&=\mu_{\rm Dec}\left(t-t_{\rm ej, decel}\right)-\frac{1}{2} \dot{\mu}_{\rm Dec}\left(t-t_{\rm ej, decel}\right)^{2},
\end{split}
\label{eq:decel}
\end{equation}
where variables are the same as those in Equation~\ref{eq:bulkmotion}, except $\dot{\mu}_{\rm R.A.}$ and $\dot{\mu}_{\rm Dec}$ are the R.A. and Declination acceleration terms (in units of mas\,day$^{-2}$). Using Equation~\ref{eq:decel}, the time of zero separation is estimated to be $t_{\rm ej, decel}$=MJD~58024.1$^{+2.6}_{-3.2}$. 

In the third case, we use a combination of the previous two models. Here, S2 can initially travel with a constant motion (described by Equation~\ref{eq:bulkmotion}) until time = $t_{\rm decel}$, following which, S2 can decelerate (Equation~\ref{eq:decel}). This estimates the time of ejection to be $t_{\rm ej, comb}$=MJD~58010.8$^{+2.65}_{-2.5}$ and the time of deceleration to occur at $t_{\rm decel}$=MJD~58262$^{+32}_{-65}$.

\begin{table*}
\caption{MCMC modelling of the proper motion of S2. Here we use a bulk motion model (Equation~\ref{eq:bulkmotion}), a decelerating motion model (Equation~\ref{eq:decel}), or a combination of the two. For both R.A. and Dec, we show the average proper motion as $\bar{\mu}_\alpha$ and $\bar{\mu}_\delta$, respectively. We also show the average acceleration in both R.A. and Dec ($\dot{\mu}_\alpha$ and $\dot{\mu}_\delta$, respectively), the average proper motion $\mu_{\rm ave}$, the deceleration start date $t_{\rm decel}$ for the combination model), as well as the best-fit date of the ejection ($t_{\rm ej}$) in MJD.}
\centering
\label{tab:ejecta_modelling}
\begin{tabular}{ccccccccc}
\hline
        & $\bar{\mu}_{\alpha}$  & $\bar{\mu}_{\delta}$ & $t_{\rm decel}$ & $\dot{\mu}_{\alpha}$ & $\dot{\mu}_{\delta}$ & $\mu_{\rm ave}$ & $t_{\rm ej}$ \\
        & (mas\,day$^{-1}$) & (mas\,day$^{-1}$) & (MJD)  & (mas\,day$^{-2}$) & (mas\,day$^{-2}$) & (mas\,day$^{-1}$) & (MJD) \\
        \hline
Bulk motion & 36.4$\pm$0.4 & -25.35$^{+0.25}_{-0.31}$ & --  & -- & -- & 44.37$^{+0.6}_{-0.8}$ & 58003.4$^{+1.6}_{-1.7}$\\   

Decelerating motion & 46.57$^{+1.67}_{-1.83}$ & -34.0$^{+1.2}_{-1.3}$ & -- & 0.05$\pm$0.01 & -0.05$\pm$0.01 & 57.6$\pm$3.0  & 58024.1$^{+2.6}_{-3.2}$ \\

Combination (bulk+decel) & 38.7$^{+1.0}_{-0.7}$ & -27.0$^{+0.4}_{-0.7}$ & 58262$^{+32}_{-65}$  & 0.11$^{+0.08}_{-0.04}$ & -0.13$^{+0.06}_{-0.09}$ & 47.2$\pm$1.5 & 58010.8$^{+2.65}_{-2.5}$\\

\hline
\end{tabular}
\end{table*}

\begin{figure}[!ht]
\centering
\includegraphics[width=1\columnwidth]{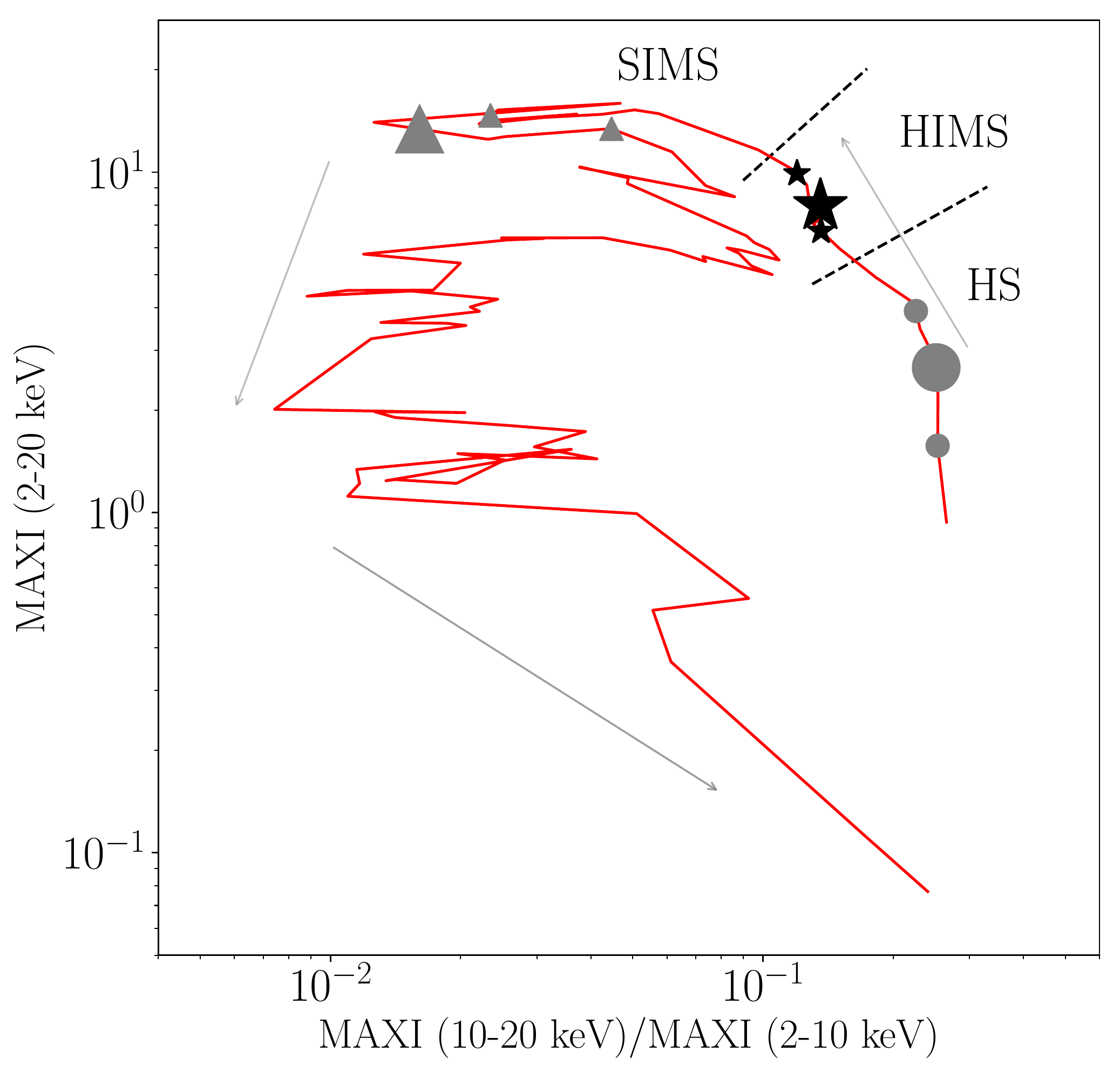}
\caption{HID of \source{} during its major outburst (see Figure~\ref{fig:HID} for full details). Here, the black stars represent the best fit ejection time, assuming S2 initially travelled with a constant motion before decelerating. The largest marker represents the best fit, while the smaller show the extent of the 1-$\sigma$ errors. We also show the estimated ejection times when we assume constant motion (grey circles), as well as allowing S2 to decelerate at all times (grey triangles). We mark the state transitions during the rise of the outburst (dashed lines) and the arrows indicate how the source evolved with time. The determined timing of the ejection event in comparison to the source evolution implies that the constant motion plus deceleration model best describes the data.}
\label{fig:HID_ejecta}
\end{figure}

We then compare these results (from Table~\ref{tab:ejecta_modelling}) with the X-ray and radio monitoring to determine the most plausible scenario (Figure~\ref{fig:lc}). Assuming simple bulk motion implies $t_{\rm ej}$ would have occurred early in the outburst, during the hard state when the source was brightening (Figure~\ref{fig:HID_ejecta}). This ejection window is well before the expected time for a jet ejection, which is typically believed to occur close in time to the transition to the SIMS or soft state  \citep[e.g.,][]{2004MNRAS.355.1105F,2004ApJ...617.1272C,2009MNRAS.396.1370F,2012MNRAS.419L..49M}. Additionally, over this time we observe the brightening and then steady emission from the compact jet. The launching window is also $\approx$10\,days prior to the onset of jet quenching in the radio band, with compact emission detected in the IR \citep{2018ApJ...867..114B}, and mm \citep{2017ATel10745....1T} indicating the presence of the compact jet even at high energies (closer to the BH). 

When deceleration was included for all times since the ejection (Equation~\ref{eq:decel}), $t_{\rm ej, decel}$ was estimated to occur at a time when the X-ray emission was fading, and the source was undergoing spectral hardening (as it moved back towards a hard state; Figure~\ref{fig:HID_ejecta}). Additionally, this ejection date is $>$4 days after the end of the initial radio flare (Figure~\ref{fig:variability}), and $>$1.6\,days after we see rising radio emission indicating the second radio flare had already started (Figure~\ref{fig:variability_58019}). Given that a radio flare is caused by the ejected material moving away from the BH (to distances probed by the radio band), the flares should occur after the ejection event and not before. Hence, we also deem this scenario to be improbable.

Therefore, the most plausible scenario is the combination of the two models, where S2 travelled with an $\approx$constant velocity for the first $\sim$260\,days, before slowing as it interacted with a denser region of the jet or surrounding environment, or once it had swept up enough ISM to be equal to its own mass. This model places the launching time close to the HIMS to SIMS transition (see Section~\ref{sec:ejecta_connection} for further discussion).

\section{Discussion}

\subsection{The evolving radio jet}

Our multi-frequency radio observations of \source{} show the evolution of the jet throughout its 2017/2018 major outburst. These observations probed the initial brightening of the compact radio jet during the hard state, the subsequent quenching of the compact jet, and radio flaring as the source moved through the intermediate states into the soft state. Our ATCA monitoring detected the re-appearance of the compact jet during a short-lived return to the hard state. We did not detect radio emission from the jet as it re-established over the hard-to-soft return state transition at the end of this major outburst.

Additionally, over a period of nearly a year, ATCA and MeerKAT observations traced a spatially-resolved downstream jet knot S2 as it moved away from the black hole, allowing for constraints on the properties of the jet at the time of the ejection.

\subsubsection{Hard state radio/X-ray correlation}
\label{sec:lrlx}
In their hard states, BH XRBs exhibit an empirical correlation between their radio ($L_{\rm R}$) and X-ray ($L_{\rm X}$) luminosities, which is observed over several orders of magnitude in luminosity \citep[e.g.,][]{2000A&A...359..251C,2003A&A...400.1007C,2003MNRAS.344...60G,2012MNRAS.423..590G,2013MNRAS.428.2500C,2018MNRAS.478L.132G}. This non-linear relationship is generally described by two power-law tracks: an upper `radio-loud' track with a slope of $L_{\rm R} \propto L_{\rm X}^{\sim 0.6}$, and a lower `radio-quiet' track with a slope of $L_{\rm R} \propto L_{\rm X}^{\sim 1}$ \citep[e.g.,][]{2011MNRAS.414..677C,2012MNRAS.423..590G}, which show different radio spectral indices \citep{2018MNRAS.473.4122E}. We note that recent results have questioned the statistical significance of there being two separate tracks \citep{2014MNRAS.445..290G,2018MNRAS.478L.132G}.

\begin{figure}[!t]
\centering
\includegraphics[width=0.99\columnwidth]{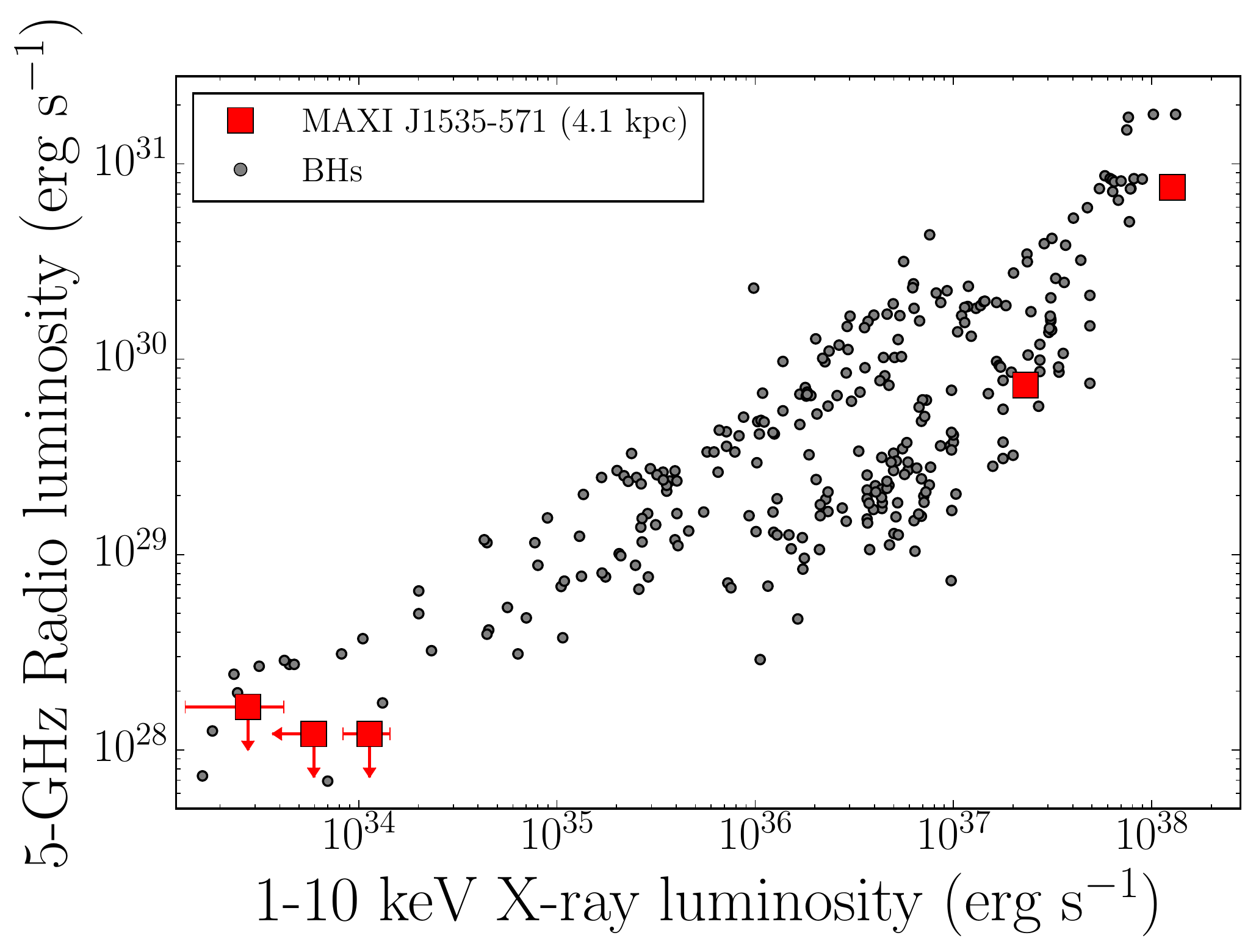}
\caption{Hard state radio and X-ray measurements of \source{} during its major outburst (of which there were only two quasi-simultaneous hard state detections). We also show the (3-$\sigma$) radio non-detections during the reverse transition at the end of the outburst. Here, the red squares show the luminosities for the estimated source distance of 4.1\,kpc \citep{2019arXiv190508497C}. The larger sample of BH systems are shown (from \citealt{2013MNRAS.428.2500C,arash_bahramian_2018_1252036}). Our two contemporaneous hard state radio and X-ray detections suggest that \source{} lies on the radio-quiet track of the radio/X-ray correlation.}
\label{fig:lrlx}
\end{figure}

We investigated the radio and X-ray relationship of \source{} by placing contemporaneous hard state 5\,GHz radio and 1-10\,keV X-ray luminosities on the radio/X-ray plane (Figure~\ref{fig:lrlx}). Unfortunately, due to the low cadence of our radio observations during the rising hard state, and the non-detection of radio emission following the transition back to the hard state at the end of the major outburst, we are only able to place two detections on the radio/X-ray correlation (with the upper limits showing the non-detection of the jet at the end of the outburst). From these two points, we determine a slope of $L_{\rm R} \propto L_{\rm X}^{1.37\pm0.05}$. However, while it appears that \source{} traced out the radio-quiet track, these two points only span 1 order of magnitude in $L_{\rm R}$ and a factor of $\sim$6 in $L_{\rm X}$, where previous studies \citep[e.g.][]{2013MNRAS.428.2500C} have shown significant deviation from a source's standard behaviour for luminosity ranges $<$2 orders of magnitude.

\subsubsection{Quenching of the compact jet}
Following a period of relatively steady radio emission from the compact jet ($\sim$MJD~58008 until MJD~58012; Figure~\ref{fig:lc}), we observed the initial stages of jet quenching. ATCA radio observations on MJD~58013.5 showed that the radio emission had faded by a $\gtrsim$25\% (and was continuing to fade) and the jet spectral break had evolved into the radio band (Section~\ref{sec:lightcurves}, but will be discussed in detail by Russell et al.\ in preparation), driving the observed jet quenching \citep[e.g.][]{1999MNRAS.304..865F,2011MNRAS.414..677C,2013MNRAS.436.2625V,2013ApJ...768L..35R,2014MNRAS.439.1390R}.

The compact jet also re-formed during \source's brief return to the hard state, quenching once again as the source transitioned back to the soft state (through the intermediate states). This second quenching was also characterised by fading radio emission and a steepening radio spectrum. 

The radio emission from \source{} faded below our detection limits on $\sim$MJD~58166. Comparing our deepest soft-state radio upper-limit with the steady, flat-spectrum radio emission measured during the HIMS provides a lower-limit of $>$3.5 orders of magnitude on the jet quenching factor. This is the deepest constraint on the soft state jet quenching to date, suggesting that the compact jet was not present during this soft state \citep[see also][]{1999MNRAS.304..865F,2011MNRAS.414..677C,2011ApJ...739L..19R}. However, our observations do not rule out jets with low-radiative efficiency in the soft states \citep{2005ApJ...625...72S,2017MNRAS.466.4272D}, although see \citet{2018A&A...612A..27K} for evidence against a dark jet during the soft state in Cygnus~X-3.

\subsubsection{Reverse transition at the end of the outburst}

At the end of its major outburst, \source{} transitioned from the soft to the hard state. This reverse transition was interesting for two reasons: 1) it occurred at a much lower X-ray luminosity than is generally expected, and; 2) the radio jet was not detected.

In a typical BH XRB outburst, the hard-to-HIMS transition during the outburst rise occurs at an X-ray luminosity of $\geq$3\%\,\lEdd, while the lower-luminosity transition back to the hard state at the end of an outburst occurs between 0.3\% and 10\% of $L_{\rm Edd}$ \citep{2010MNRAS.403...61D,2013ApJ...779...95K,2019MNRAS.485.2744V}, with an average value of $\sim$2\% \citep{2003A&A...409..697M}. However, \source{} only transitioned back to the hard state at $L_{\rm X} \sim 0.003$\%\,\lEdd{} (see also \citealt{2019arXiv190508497C}). Such low-luminosity transitions are atypical for BH XRBs, and the only other source to show similar behaviour is 4U~1630$-$47, which transitioned from the soft to hard state at $L_{\rm X} \approx$0.008\%\,\lEdd{} \citep{2014ApJ...791...70T}. 

Low transition luminosities have been attributed to either the decay being disrupted by a new mass inflow re-igniting and extending the soft state to a lower than usual X-ray luminosity \citep{2019MNRAS.485.2744V}, or due to a low disk magnetic field and viscosity \citep[e.g.,][]{2008MNRAS.385L..88P,2014ApJ...782L..18B}. 

For the case of a new mass inflow extending the soft state and pushing the reverse transition to much lower than typical luminosities, the jet may have been undetected due to the low transition luminosity. Assuming typical hard-state \lrlx{} scalings (Figure~\ref{fig:lrlx}), the X-ray luminosity over the reverse transition at the end of \source's outburst implies an expected radio flux density of $\sim$50--400\,\uJy. Therefore, while our 3-$\sigma$ radio upper-limits are relatively radio-faint in comparison with the majority of other hard state black holes, the radio non-detection is not remarkable. 

In the case where low disk magnetic fields were responsible for the low-luminosity state transition, we may also expect weaker radio jets \citep[e.g.,][]{1986PASJ...38..631S,2012A&A...538A...5K,2014ApJ...782L..18B,2015A&A...574A.133K}, resulting in the radio non-detection. Our data do not allow us to conclusively determine the cause of the low-luminosity transition or non-detection of the radio jets.

\subsection{S2, an apparently superluminal jet knot}
\label{sec:S2_discussion}

Our radio monitoring tracked the motion of the jet knot, S2, as it moved away from the core position of \source{} (Figures~\ref{fig:ejectaimage} and \ref{fig:ejecta_motion}). We only detected a single-sided ejection. Assuming a bi-polar ejection \citep[e.g.,][]{1994Natur.371...46M,1999MNRAS.304..865F}, the one-sided detection could be due to S2 being the approaching component, hence, the non-detection of a receding component could be due to Doppler boosting effects reducing the flux density as it receded 
\citep[for details see e.g.,][]{2001ApJ...553..766M,2011MNRAS.415..306M}. Possible alternatives to explain the non-detection of a counter-jet component include absorption effects, lack of internal and external shocks (within the jet, or with the surrounding medium), optical depth effects, or asymmetries in jet launching \citep[e.g.,][]{1995Natur.375..464H,2013ApJ...774...12F}.

Following its initial detection on MJD~58090, S2 was detected in all radio observations until MJD~58139 (over 40\,days), when it dropped below our detection threshold. During these detections, S2 faded steadily (Figure~\ref{fig:lc}), likely as it expanded adiabatically. However, S2 also re-brightened at later times, being detected during a MeerKAT observation on MJD~58222, as well as consecutive ATCA observations on MJDs~58252 and 58255. S2 was again detected much later, during an ATCA observation on MJD~58393, which was 303\,days after its initial detection. While S2 was not detected during our monitoring on other dates before, between, and after these re-brightenings (see Table~\ref{tab:ATCA_data_S2}). As highlighted by the MeerKAT detection (at 1.3\,GHz, which would translate to a 5.5\,GHz flux density of $\sim$105\,$\mu$Jy) it is possible that S2 was below the ATCA detection threshold for some of these observations.

The re-brightenings of S2 at these later times could be produced by internal shocks with the pre-existing jet or interactions with the ISM, where inhomogentities can cause the multiple brightenings \citep[e.g.,][]{2002Sci...298..196C,2003ApJ...582..945K,2005ApJ...632..504C,2017MNRAS.472..141M}. Such interactions with the ISM could also lead to S2 decelerating at later times. Alternatively, the slowing may also have occurred once the S2 had swept up enough ISM material (equal to its own mass) to slow \citep[e.g.,][]{2002Sci...298..196C}.

\subsubsection{Jet knot properties: speed, inclination, opening angle, expansion, and energetics}

The proper motion of 47.2$\pm$1.5\,mas\,day$^{-1}$ indicates an apparent jet velocity of $\approx$1.1c for a source distance of 4.1\,kpc. Apparent superluminal motions of jet ejecta have been observed in a handful of BH XRBs to date (e.g., \citealt{1994Natur.371...46M,1995Natur.374..141T,1995Natur.375..464H,1998IAUC.6938....2R,1999ARA&A..37..409M,1999MNRAS.304..865F,1999ApJ...511..398R,2002MNRAS.336...39F,2002Sci...298..196C}, see \citealt{2006csxs.book..381F}, for review), and indicates that S2 was the approaching component.

\begin{figure}[!t]
\centering
\includegraphics[width=0.99\columnwidth]{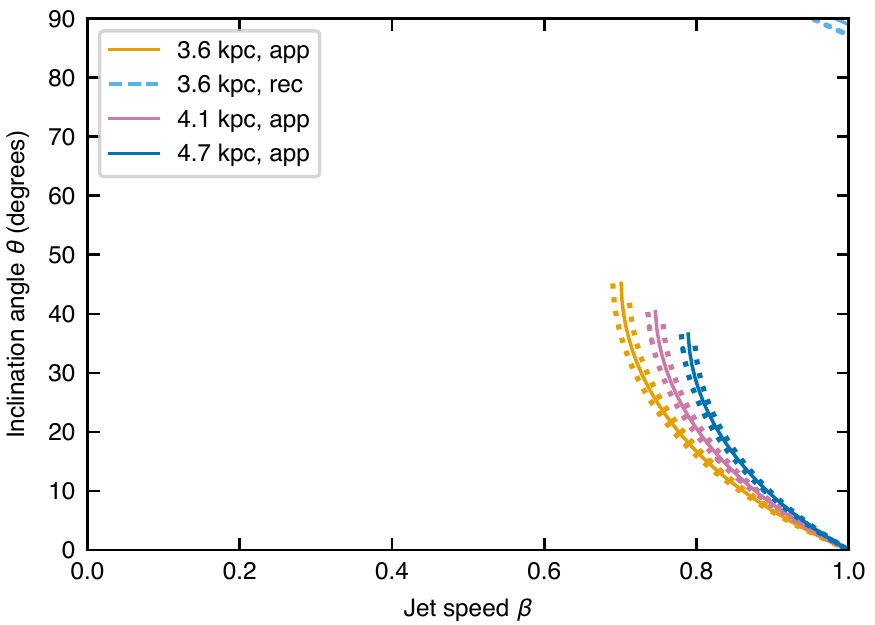}
\caption{Constraints on the jet speed and inclination angle to the line of sight from the proper motion for the full range of source distances (4.1$^{+0.6}_{-0.5}$\,kpc) presented by \citet{2019arXiv190508497C}. Uncertainties are shown as the dotted lines. For all but the lowest distance limit (of 3.6\,kpc) S2 must be the approaching component. }
\label{fig:inclination_speed}
\end{figure}

Our tracking of S2 can be used to constrain the inclination of the jet (at the time of the ejection event; \citealt{2019Natur.569..374M}) and the speed of the ejection. We did not detect a counter-jet component, so we cannot uniquely solve for the jet speed, $\beta = \frac{v}{c}$, or inclination, $\theta$. Instead, we can only solve for $\beta \cos{\theta}$ , given that \citep[e.g.,][]{1966Natur.211..468R,1994Natur.371...46M}:
\begin{equation}
\mu_{\substack{{\rm app}\\{\rm rec}}} = \frac{\beta \sin{\theta}}{1 \mp \beta \cos{\theta}}\frac{c}{D},
\label{eq:bcost}
\end{equation}
\noindent where $\mu_{\rm app}$ and $\mu_{\rm rec}$ are the approaching and receding proper motions, and $D$ is the distance to the source. As shown in Figure~\ref{fig:inclination_speed}, from its apparent superluminal motion, S2 is almost certainly the approaching component, with only a small set of solutions existing for it to be the receding component (only at the lowest distance limit; \citealt{2019arXiv190508497C}), $\beta \cos{\theta} \geq 0.49$, such that $\beta \geq 0.69$ and $\theta \leq 45^{\circ}$.

X-ray spectral fits of the iron line during the HIMS \citep{2018ApJ...852L..34X,2018ApJ...860L..28M} favoured a disk inclination of $\sim$55--68$^\circ$, which is discrepant from our determined values for S2. However, there was some evidence for disk warping \citep{2018ApJ...860L..28M}, suggesting that the inner disk orientation was changing over time. Such changes in the inner disk and jet orientation can be rapid \citep[e.g.,][]{2018MNRAS.474L..81L,2019Natur.569..374M}, and would account for such a difference between the disk inclination and the jet when measured at different times. Additionally, the jet and outer disk may be misaligned.

S2 remained unresolved for all radio detections. Therefore, while we do not observe the expansion of S2, our observations can constrain the opening angle and transverse expansion of the jet from its width at a given distance from the core \citep[e.g.,][]{1999MNRAS.304..865F}. Combining the 2.9$^{\prime\prime}$ ATCA resolution with the $\sim$17$^{\prime\prime}$ separation we measured when S2 was detected at its maximum separation, the jet opening angle is constrained to $\leq$10$^\circ$, similar to typical constraints on BH XRB jet opening angles \citep[see e.g.,][]{2006MNRAS.367.1432M}. In terms of the transverse expansion, combining the ATCA beamsize of our final detection with the ejection date, we limit the expansion velocity to $\leq$0.18c, consistent with the expansion estimates found for V404~Cygni \citep{2017MNRAS.469.3141T}.

Using the proper motion of S2, we also estimate the size scale of the radio emission based on the delay between the ejection time and the time of the first radio flare. While we do not detect the peak of the first radio flare, we constrain it to have occurred between MJDs~58013.6 and 58017.4 (Section~\ref{sec:radio_core}). Therefore, the $\sim$GHz radio emitting region lies at a distance of $<$430\,mas from the radio core. For a source distance of 4.1\,kpc, this corresponds to a size of $<$1760\,AU. 

From the rise time and brightness reached by the second radio flare\footnote{We did not adequately sample the first radio flare with our radio monitoring to estimate the radio brightness and rise time.}, we also place constraints on the minimum energy and magnetic field required to produce such a flare. Following \citet{2006csxs.book..381F} and assuming equipartition between electrons and magnetic field, and one proton per electron, we estimate the minimum energy $E_{\rm min}  \sim 10^{42}$--$10^{43}$\,erg, corresponding to a minimum mean power $P_{\rm min} \sim 10^{37}$\,erg\,s$^{-1}$ with an equipartition magnetic field, $B_{\rm eq} \sim$10--500 mGauss. These values are comparable to estimates from observed ejection events from a number of other sources \citep[e.g.,][]{1999MNRAS.304..865F,2007MNRAS.378.1111B,2014MNRAS.437.3265C}. Our energy estimates differ from the ejecta observed from V404~Cygni \citep{2017MNRAS.469.3141T,2019Natur.569..374M}, however, in that outburst, multiple, smaller ejecta were observed, explaining the lower energies and higher magnetic fields.

\subsubsection{X-ray properties at the time of the ejections}
\label{sec:ejecta_connection}

While there is a connection between the morphology of the jets and structure of the accretion flow \citep[e.g.][]{2004MNRAS.355.1105F}, the causal sequence of events leading to the changes in the jets is not well understood, and the coupling is likely to be complex. The onset of the transient jet is generally linked to the transition from the HIMS to the SIMS. The X-ray properties over this transition show a marked change \citep[e.g.][]{2005A&A...440..207B}. For example, there is a sharp decrease in the fractional rms variability of the X-ray emission and the sudden appearance of a Type-B QPO. It is these changes in the X-ray emission that have often been linked to the moment of an ejection event, in particular the presence of the type-B QPO \citep[e.g.][]{2006csxs.book..381F,2008ApJ...675.1407K,2009MNRAS.396.1370F,2012MNRAS.419L..49M}. However, sparse observational sampling and the delay between an ejection and the observed radio flaring (due to optical depth effects and the time required for shocks to occur) generally prevents such a connection being clearly identified \citep[see][for a review]{2009MNRAS.396.1370F}. 

During our radio coverage of the outburst of \source, we observed the onset of the compact jet quenching in the radio band (which started $\sim$MJD~58013.6). Our next radio observation (starting on MJD~58017.4) detected the end of a radio flare. Therefore, this initial radio flare occurred between MJDs\,58013.6 and 58017.4 and was likely associated with the ejection of S2, meaning that we would typically expect the ejection to have occurred at an earlier time (due to the time delay between an ejection event and the subsequent radio flaring; \citealt{2009MNRAS.396.1370F}).

Comparing our ejection window (between MJD~58008.3 and MJD~58013.4) to the X-ray properties produces interesting results. \source{} transitioned from the HIMS to SIMS sometime between MJD~58014.18 and MJD~58015.37 \citep{2018MNRAS.480.4443T,2018ApJ...866..122H}, close in time to the estimated ejection of S2. \citet{2018ApJ...866..122H} report a detection of a possible type-B QPO during HXMT observations taken on MJD~58015.97 (although this QPO was observed at a relatively high QPO frequency of $\sim$10\,Hz). While this QPO could be consistent with being close to the peak of the first radio flare (but certainly not before the beginning of the flare), it is after our $t_{\rm ej}$ window. In observations after this potential type-B QPO detection, only type-C QPOs were observed until MJD~58017.5. HXMT and AstroSat observations taken during our ejection event window show high fractional X-ray variability ($\gtrsim$10\%) with only the presence of type-C QPOs, with a changing QPO frequency (the QPO frequency first decreased, reaching a minimum $\sim$MJD~58010, before increasing during the ejection window, although some scatter was observed; \citealt{2018ApJ...866..122H,2019MNRAS.487..928S,2019arXiv190610595B}).

Conveniently, the Neutron Star Interior Composition Explorer (NICER) X-ray telescope on-board the International Space Station (ISS) densely monitored \source{} around the time of the HIMS to SIMS transition (observing the source multiple times nearly every day). As reported by \citet{2018ApJ...865L..15S}, the NICER X-ray observations also detected the appearance of a possible type-B QPO during observations starting on MJD~58016.8, that remained until MJD~58025 (but not during observations earlier and, in particular, at similar times to when HXMT reported a type-B QPO). While this type-B QPO could be coincident with the second radio flare, it is not consistent with $t_{\rm ej}$. Additionally, NICER's high observing cadence showed the X-ray fractional variability was relatively high (dropping from $\approx$15\% to $\approx$11\%) over our $t_{\rm ej}$ window, making it seem unlikely that a type-B QPO was present previously (typically, type-B QPOs occur at times of low fractional X-ray variability, $\approx$3--5\% rms; \citealt{2010LNP...794...53B}).

The X-ray observations allow us to further investigate the X-ray properties at the time of $t_{\rm ej}$. There was an initially steady drop in the X-ray rms variability (from $\approx$15\% to $\approx$11\% rms) between MJDs~58011 and 58014, followed by a more rapid decrease ($\approx$11\% to $\approx$7\% rms) between MJDs~58013 and 58015 \citep[see figure 1 in][]{2018ApJ...865L..15S}. Over this time, the X-ray observations also showed a steady drop in X-ray hardness, as well as an interesting and relatively sudden increase in the soft X-rays, where the count rate increased rapidly by a factor of $\sim$2 between MJDs~58010 and 58014 \citep[see figure 1 in][]{2018ApJ...865L..15S}. While this increase could be due to a change in the accretion properties, similar short high-energy brightenings are commonly associated to ejection events in AGN \citep[e.g.,][]{2008Natur.452..966M,2016NatPh..12..772K,2017MNRAS.468.4478L}, where the X-ray increase can arise from inverse Compton scattering of the synchrotron radiation from the knot after it was ejected. If this X-ray increase was indeed related to the ejection of S2, it favours an ejection time of $\sim$MJD~58010. Additionally, around the same time the QPO frequency rapidly decreased, reaching a local minimum $\sim$MJD~58010 (after which it increased again; \citealt{2018ApJ...866..122H,2019MNRAS.487..928S}). Speculatively, if the QPO frequency is related to the radius of the X-ray emitting material \citep[e.g.,][]{2009MNRAS.397L.101I}, we may expect the QPO frequency to decrease due the extraction of accreting material in a jet ejection \citep{2014MNRAS.440.2882R}. In addition, there was also a change in the QPO behaviour at around MJD~58013, when the QPO frequency varied rapidly \citep{2019arXiv190610595B}.

Using high-resolution radio observations of the 2009 outburst of H1743$-$322, \citet{2012MNRAS.421..468M} were able track the motion of a bipolar ejection over two epochs, allowing tight constraints on the time of ejection (to within 1\,day). While their estimated ejection date unfortunately coincided with a 3\,day gap in X-ray monitoring, they determined that it occurred immediately prior to the HIMS $\rightarrow$ SIMS transition. Over this time, H1743$-$322 displayed a short increase (also by a factor of $\sim$2) in the X-ray count rate, a rapid reduction in the X-ray rms variability, an evolution in the type-C QPOs, and the onset of compact jet quenching. Additionally, \citet{2012MNRAS.421..468M} only report the first appearance of a type-B QPO $\sim$4\,days after the estimated ejection event. 

The similarities shown by H1743$-$322 and \source{} are striking. While in both cases we are unable to conclusively rule out the presence of a type-B QPO at the time of the jet ejection, the results suggest that, for these two outbursts, the X-ray signature of the ejection was not the appearance of the type-B QPO. It could be that the jet ejection and type-B QPO are both a result of some other effect. Similar to findings reported by \citet{2012MNRAS.421..468M}, our monitoring implies that the ejection event was instead related to the rapid drop in X-ray rms variability immediately prior to the HIMS $\rightarrow$ SIMS transition, the sudden increase in soft X-ray count rate, or the change in the type-C QPO frequency.

\citet{2009MNRAS.396.1370F} and \citet{2012MNRAS.421..468M} also compared outburst data from a number of other BH LMXBs, using either VLBI data to trace ejecta back in time, or by connecting the timing of the radio-flares to the X-ray behaviour. Their analysis found no clear evidence of an association between the jet ejection and the appearance of type-B QPOs. In most cases, it appeared that the time of ejection was contemporaneous with a change in the type-C QPO and a decrease in the X-ray rms variability. However, as discussed by \citet{2009MNRAS.396.1370F} and \citet{2012MNRAS.421..468M}, this did not hold true for all systems, or even outbursts from the same system. Radio and X-ray observations of the 2002 outburst of GX~339$-$4 showed that while the type-C QPO was changing at the time of the radio flare \citep{2004MNRAS.347L..52G,2005A&A...440..207B}, the drop in X-ray rms variability was observed a few days after the radio flare (where the rms drop occurred at around the time of the detection of type-B QPO; \citealt{2009MNRAS.396.1370F}). Additionally, the 2003 outburst of H1743$-$322 did not appear to show an evolution of the type-C QPO during the estimated time of ejection \citep{2012MNRAS.421..468M}. 

Therefore, while our \source{} results agree well with the 2009 outburst of H1743$-$322, as well as a number of other systems (as presented by \citealt{2009MNRAS.396.1370F} and \citealt{2012MNRAS.421..468M}) comparisons with the 2002 outburst of GX~339$-$4 and 2003 outburst of H1743$-$322 muddy the picture. These two results imply that the events driving an ejection event may vary between systems and even outbursts of the same system. Alternatively, as discussed by \citet{2014SSRv..183..323F} there may be no clear X-ray signature to the moment of ejection, and the observed knots could be a result of internal shocks arising within the jets from rapid (but not instantaneous) changes in the injection, or speed of the jet-channeled accretion material \citep[e.g.,][]{2010MNRAS.401..394J,2013MNRAS.429L..20M}.

\section{Conclusions}

With our comprehensive radio monitoring of \source{} during its 2017/2018 major outburst, we have observed the evolution of the compact jet, as well as tracked the motion of a downstream jet knot. 

Our observations constrain the compact jet quenching to be a factor of $>$3.5 orders of magnitude, implying that the compact jet was not visible during the soft state. Interestingly, the radio jet was undetected by our observations during the exceptionally low X-ray luminosity reverse (soft to hard state) transition at the end of the outburst, when we expect the compact jet to re-brighten.  

From the observed radio flare and detection of the discrete, apparently superluminal jet knot, we place constraints on the properties of the jet. We estimate a jet opening angle of $<$10$^{\circ}$. We determine a jet inclination of $\leq45^{\circ}$ at the time of ejection and a jet velocity of $\beta \geq 0.69$. 

Extrapolating the motion of the knot back in time to determine the time of ejection reveals that in this outburst of \source{} the ejection likely occurred a few days before the appearance of a possible type-B QPO in X-ray monitoring (which has often been associated with the transient jet launching). Instead, our results suggest that the ejection may be linked to the short increase in X-ray count rate, the observed drop in X-ray variability, or the change in the type-C QPO frequency, which was observed immediately before HIMS to SIMS transition.

\acknowledgments
We thank Jamie Stevens and staff from the Australia Telescope National Facility (ATNF) for scheduling the ATCA radio observations. We also thank \swift{} for the scheduling of the X-ray observations. TDR thanks Abbie Stevens, Phil Uttley, and Craig Anderson for helpful discussions. TDR acknowledges support from the Netherlands Organisation for Scientific Research (NWO) Veni Fellowship, grant number 639.041.646. AJT is supported in part by an Natural Sciences and Engineering Research Council of Canada (NSERC) Post-Graduate Doctoral Scholarship (PGSD2-490318-2016). JCAM-J is the recipient of an Australian Research Council Future Fellowship (FT140101082), funded by the Australian government. GRS and AJT acknowledge support from an NSERC Discovery Grant (RGPIN-06569-2016). ASP and RW are supported by a NWO Top grant, module 1, awarded to RW. SC and ET acknowledge financial support from the UnivEarthS Labex program of Sorbonne Paris Cit\'{e} (ANR-10-LABX-0023 and ANR-11-IDEX-0005-02). DA acknowledges support from the Royal Society. ND and JvdE are supported by a NWO Vidi grant, awarded to ND. SM and ML are supported by a NWO Vici grant, awarded to SM (grant number 639.043.513). PAW acknowledges support from the NRF and UCT. This work was supported in part by the Oxford Hintze Centre for Astrophysical Surveys which is funded through generous support from the Hintze Family Charitable Foundation. The International Centre for Radio Astronomy Research is a joint venture between Curtin University and the University of Western Australia, funded by the state government of Western Australia and the joint venture partners. The Australia Telescope Compact Array is part of the Australia Telescope, which is funded by the Commonwealth of Australia for operation as a National Facility managed by CSIRO. The MeerKAT telescope is operated by the South African Radio Astronomy Observatory, which is a facility of the National Research Foundation, an agency of the Department of Science and Technology. This research has made use of NASA's Astrophysics Data System.

\vspace{5mm}
\facilities{ATCA, MeerKAT, \swift-XRT, MAXI}

\software{AOFlagger (version 2.9; \citealt{2010offringa}), Astropy; \citep{2013A&A...558A..33A}, \textsc{CASA} (version 5.1.0; \citealt{2007ASPC..376..127M}), \textsc{emcee} \citep{2013PASP..125..306F}, killMS (https://github.com/saopicc/killMS), \textsc{UVMULTIFIT} \citep{2014A&A...563A.136M}, \textsc{XSPEC} (version 12.9.1; \citealt{1996ASPC..101...17A})}

\bibliographystyle{apj}

\begin{thebibliography}{}

\bibitem[\protect\citeauthoryear{{Arnaud}}{{Arnaud}}{1996}]{1996ASPC..101...17A}
{Arnaud}, K.~A. 1996, in Astronomical Society of the Pacific Conference Series,
  Vol. 101, Astronomical Data Analysis Software and Systems V, ed. G.~H.
  {Jacoby} \& J.~{Barnes}, 17

\bibitem[\protect\citeauthoryear{{Astropy Collaboration} et~al.}{{Astropy
  Collaboration} et~al.}{2013}]{2013A&A...558A..33A}
{Astropy Collaboration}, et~al. 2013, \aap, 558, A33

\bibitem[\protect\citeauthoryear{{Baglio} et~al.}{{Baglio}
  et~al.}{2018}]{2018ApJ...867..114B}
{Baglio}, M.~C., et~al. 2018, \apj, 867, 114

\bibitem[\protect\citeauthoryear{Bahramian et~al.}{Bahramian
  et~al.}{2018}]{arash_bahramian_2018_1252036}
Bahramian, A., et~al. 2018, {Radio/X-ray correlation database for X-ray
  binaries}

\bibitem[\protect\citeauthoryear{{Barkana} \& {Loeb}}{{Barkana} \&
  {Loeb}}{2001}]{2001PhR...349..125B}
{Barkana}, R.,  \& {Loeb}, A. 2001, \physrep, 349, 125

\bibitem[\protect\citeauthoryear{{Begelman} \& {Armitage}}{{Begelman} \&
  {Armitage}}{2014}]{2014ApJ...782L..18B}
{Begelman}, M.~C.,  \& {Armitage}, P.~J. 2014, \apjl, 782, L18

\bibitem[\protect\citeauthoryear{{Belloni} et~al.}{{Belloni}
  et~al.}{2005}]{2005A&A...440..207B}
{Belloni}, T., {Homan}, J., {Casella}, P., {van der Klis}, M., {Nespoli}, E.,
  {Lewin}, W.~H.~G., {Miller}, J.~M.,  \& {M{\'e}ndez}, M. 2005, \aap, 440, 207

\bibitem[\protect\citeauthoryear{{Belloni}}{{Belloni}}{2010}]{2010LNP...794...53B}
{Belloni}, T.~M. 2010, {States and Transitions in Black Hole Binaries}, ed.
  T.~{Belloni} 53

\bibitem[\protect\citeauthoryear{{Bhargava} et~al.}{{Bhargava}
  et~al.}{2019}]{2019arXiv190610595B}
{Bhargava}, Y., {Belloni}, T., {Bhattacharya}, D.,  \& {Misra}, R. 2019, arXiv
  e-prints, arXiv:1906.10595

\bibitem[\protect\citeauthoryear{{Brocksopp} et~al.}{{Brocksopp}
  et~al.}{2007}]{2007MNRAS.378.1111B}
{Brocksopp}, C., {Miller-Jones}, J.~C.~A., {Fender}, R.~P.,  \& {Stappers},
  B.~W. 2007, \mnras, 378, 1111

\bibitem[\protect\citeauthoryear{{Camilo} et~al.}{{Camilo}
  et~al.}{2018}]{2018Camilo}
{Camilo}, F., et~al. 2018, \apj, 856, 180

\bibitem[\protect\citeauthoryear{{Casella}, {Belloni}, \& {Stella}}{{Casella}
  et~al.}{2005}]{2005ApJ...629..403C}
{Casella}, P., {Belloni}, T.,  \& {Stella}, L. 2005, \apj, 629, 403

\bibitem[\protect\citeauthoryear{{Ceccobello} et~al.}{{Ceccobello}
  et~al.}{2018}]{2018MNRAS.473.4417C}
{Ceccobello}, C., {Cavecchi}, Y., {Heemskerk}, M.~H.~M., {Markoff}, S.,
  {Polko}, P.,  \& {Meier}, D. 2018, \mnras, 473, 4417

\bibitem[\protect\citeauthoryear{{Chauhan} et~al.}{{Chauhan}
  et~al.}{2019}]{2019arXiv190508497C}
{Chauhan}, J., et~al. 2019, arXiv e-prints, arXiv:1905.08497

\bibitem[\protect\citeauthoryear{{Corbel} et~al.}{{Corbel}
  et~al.}{2013a}]{2013MNRAS.431L.107C}
{Corbel}, S., et~al. 2013a, \mnras, 431, L107

\bibitem[\protect\citeauthoryear{{Corbel} et~al.}{{Corbel}
  et~al.}{2013b}]{2013MNRAS.428.2500C}
{Corbel}, S., {Coriat}, M., {Brocksopp}, C., {Tzioumis}, A.~K., {Fender},
  R.~P., {Tomsick}, J.~A., {Buxton}, M.~M.,  \& {Bailyn}, C.~D. 2013b, \mnras,
  428, 2500

\bibitem[\protect\citeauthoryear{{Corbel} \& {Fender}}{{Corbel} \&
  {Fender}}{2002}]{2002ApJ...573L..35C}
{Corbel}, S.,  \& {Fender}, R.~P. 2002, \apjl, 573, L35

\bibitem[\protect\citeauthoryear{{Corbel} et~al.}{{Corbel}
  et~al.}{2004}]{2004ApJ...617.1272C}
{Corbel}, S., {Fender}, R.~P., {Tomsick}, J.~A., {Tzioumis}, A.~K.,  \&
  {Tingay}, S. 2004, \apj, 617, 1272

\bibitem[\protect\citeauthoryear{{Corbel} et~al.}{{Corbel}
  et~al.}{2000}]{2000A&A...359..251C}
{Corbel}, S., {Fender}, R.~P., {Tzioumis}, A.~K., {Nowak}, M., {McIntyre}, V.,
  {Durouchoux}, P.,  \& {Sood}, R. 2000, \aap, 359, 251

\bibitem[\protect\citeauthoryear{{Corbel} et~al.}{{Corbel}
  et~al.}{2002}]{2002Sci...298..196C}
{Corbel}, S., {Fender}, R.~P., {Tzioumis}, A.~K., {Tomsick}, J.~A., {Orosz},
  J.~A., {Miller}, J.~M., {Wijnands}, R.,  \& {Kaaret}, P. 2002, Science, 298,
  196

\bibitem[\protect\citeauthoryear{{Corbel} et~al.}{{Corbel}
  et~al.}{2005}]{2005ApJ...632..504C}
{Corbel}, S., {Kaaret}, P., {Fender}, R.~P., {Tzioumis}, A.~K., {Tomsick},
  J.~A.,  \& {Orosz}, J.~A. 2005, \apj, 632, 504

\bibitem[\protect\citeauthoryear{{Corbel} et~al.}{{Corbel}
  et~al.}{2003}]{2003A&A...400.1007C}
{Corbel}, S., {Nowak}, M.~A., {Fender}, R.~P., {Tzioumis}, A.~K.,  \&
  {Markoff}, S. 2003, \aap, 400, 1007

\bibitem[\protect\citeauthoryear{{Coriat} et~al.}{{Coriat}
  et~al.}{2011}]{2011MNRAS.414..677C}
{Coriat}, M., et~al. 2011, \mnras, 414, 677

\bibitem[\protect\citeauthoryear{{Curran} et~al.}{{Curran}
  et~al.}{2014}]{2014MNRAS.437.3265C}
{Curran}, P.~A., et~al. 2014, \mnras, 437, 3265

\bibitem[\protect\citeauthoryear{{Dhawan}, {Mirabel}, \&
  {Rodr{\'{\i}}guez}}{{Dhawan} et~al.}{2000}]{2000ApJ...543..373D}
{Dhawan}, V., {Mirabel}, I.~F.,  \& {Rodr{\'{\i}}guez}, L.~F. 2000, \apj, 543,
  373

\bibitem[\protect\citeauthoryear{{Dincer}}{{Dincer}}{2017}]{2017ATel10716....1D}
{Dincer}, T. 2017, The Astronomer's Telegram, 10716

\bibitem[\protect\citeauthoryear{{Drappeau} et~al.}{{Drappeau}
  et~al.}{2017}]{2017MNRAS.466.4272D}
{Drappeau}, S., et~al. 2017, \mnras, 466, 4272

\bibitem[\protect\citeauthoryear{{Dunn} et~al.}{{Dunn}
  et~al.}{2010}]{2010MNRAS.403...61D}
{Dunn}, R.~J.~H., {Fender}, R.~P., {K{\"o}rding}, E.~G., {Belloni}, T.,  \&
  {Cabanac}, C. 2010, \mnras, 403, 61

\bibitem[\protect\citeauthoryear{{Espinasse} \& {Fender}}{{Espinasse} \&
  {Fender}}{2018}]{2018MNRAS.473.4122E}
{Espinasse}, M.,  \& {Fender}, R. 2018, \mnras, 473, 4122

\bibitem[\protect\citeauthoryear{{Fabian}}{{Fabian}}{2012}]{2012ARA&A..50..455F}
{Fabian}, A.~C. 2012, \araa, 50, 455

\bibitem[\protect\citeauthoryear{{Fender}}{{Fender}}{2006}]{2006csxs.book..381F}
{Fender}, R. 2006, {Jets from X-ray binaries}, ed. W.~H.~G. {Lewin} \& M.~{van
  der Klis} (Cambridge University Press, Cambridge), 381

\bibitem[\protect\citeauthoryear{{Fender} \& {Gallo}}{{Fender} \&
  {Gallo}}{2014}]{2014SSRv..183..323F}
{Fender}, R.,  \& {Gallo}, E. 2014, \ssr, 183, 323

\bibitem[\protect\citeauthoryear{{Fender} et~al.}{{Fender}
  et~al.}{2017}]{2017fender}
{Fender}, R., et~al. 2017, arXiv e-prints, arXiv:1711.04132

\bibitem[\protect\citeauthoryear{{Fender}}{{Fender}}{2001}]{2001MNRAS.322...31F}
{Fender}, R.~P. 2001, \mnras, 322, 31

\bibitem[\protect\citeauthoryear{{Fender}, {Belloni}, \& {Gallo}}{{Fender}
  et~al.}{2004}]{2004MNRAS.355.1105F}
{Fender}, R.~P., {Belloni}, T.~M.,  \& {Gallo}, E. 2004, \mnras, 355, 1105

\bibitem[\protect\citeauthoryear{{Fender} et~al.}{{Fender}
  et~al.}{1999}]{1999MNRAS.304..865F}
{Fender}, R.~P., {Garrington}, S.~T., {McKay}, D.~J., {Muxlow}, T.~W.~B.,
  {Pooley}, G.~G., {Spencer}, R.~E., {Stirling}, A.~M.,  \& {Waltman}, E.~B.
  1999, \mnras, 304, 865

\bibitem[\protect\citeauthoryear{{Fender}, {Homan}, \& {Belloni}}{{Fender}
  et~al.}{2009}]{2009MNRAS.396.1370F}
{Fender}, R.~P., {Homan}, J.,  \& {Belloni}, T.~M. 2009, \mnras, 396, 1370

\bibitem[\protect\citeauthoryear{{Fender} et~al.}{{Fender}
  et~al.}{2002}]{2002MNRAS.336...39F}
{Fender}, R.~P., {Rayner}, D., {McCormick}, D.~G., {Muxlow}, T.~W.~B.,
  {Pooley}, G.~G., {Sault}, R.~J.,  \& {Spencer}, R.~E. 2002, \mnras, 336, 39

\bibitem[\protect\citeauthoryear{{Fendt} \& {Sheikhnezami}}{{Fendt} \&
  {Sheikhnezami}}{2013}]{2013ApJ...774...12F}
{Fendt}, C.,  \& {Sheikhnezami}, S. 2013, \apj, 774, 12

\bibitem[\protect\citeauthoryear{{Foreman-Mackey} et~al.}{{Foreman-Mackey}
  et~al.}{2013}]{2013PASP..125..306F}
{Foreman-Mackey}, D., {Hogg}, D.~W., {Lang}, D.,  \& {Goodman}, J. 2013, \pasp,
  125, 306

\bibitem[\protect\citeauthoryear{{Gallo} et~al.}{{Gallo}
  et~al.}{2004}]{2004MNRAS.347L..52G}
{Gallo}, E., {Corbel}, S., {Fender}, R.~P., {Maccarone}, T.~J.,  \& {Tzioumis},
  A.~K. 2004, \mnras, 347, L52

\bibitem[\protect\citeauthoryear{{Gallo}, {Degenaar}, \& {van den
  Eijnden}}{{Gallo} et~al.}{2018}]{2018MNRAS.478L.132G}
{Gallo}, E., {Degenaar}, N.,  \& {van den Eijnden}, J. 2018, \mnras, 478, L132

\bibitem[\protect\citeauthoryear{{Gallo} et~al.}{{Gallo}
  et~al.}{2005}]{2005Natur.436..819G}
{Gallo}, E., {Fender}, R., {Kaiser}, C., {Russell}, D., {Morganti}, R.,
  {Oosterloo}, T.,  \& {Heinz}, S. 2005, \nat, 436, 819

\bibitem[\protect\citeauthoryear{{Gallo}, {Fender}, \& {Pooley}}{{Gallo}
  et~al.}{2003}]{2003MNRAS.344...60G}
{Gallo}, E., {Fender}, R.~P.,  \& {Pooley}, G.~G. 2003, \mnras, 344, 60

\bibitem[\protect\citeauthoryear{{Gallo}, {Miller}, \& {Fender}}{{Gallo}
  et~al.}{2012}]{2012MNRAS.423..590G}
{Gallo}, E., {Miller}, B.~P.,  \& {Fender}, R. 2012, \mnras, 423, 590

\bibitem[\protect\citeauthoryear{{Gallo} et~al.}{{Gallo}
  et~al.}{2014}]{2014MNRAS.445..290G}
{Gallo}, E., et~al. 2014, \mnras, 445, 290

\bibitem[\protect\citeauthoryear{{Hjellming} \& {Rupen}}{{Hjellming} \&
  {Rupen}}{1995}]{1995Natur.375..464H}
{Hjellming}, R.~M.,  \& {Rupen}, M.~P. 1995, \nat, 375, 464

\bibitem[\protect\citeauthoryear{{Homan} et~al.}{{Homan}
  et~al.}{2001}]{2001ApJS..132..377H}
{Homan}, J., {Wijnands}, R., {van der Klis}, M., {Belloni}, T., {van Paradijs},
  J., {Klein-Wolt}, M., {Fender}, R.,  \& {M{\'e}ndez}, M. 2001, \apjs, 132,
  377

\bibitem[\protect\citeauthoryear{{Huang} et~al.}{{Huang}
  et~al.}{2018}]{2018ApJ...866..122H}
{Huang}, Y., et~al. 2018, \apj, 866, 122

\bibitem[\protect\citeauthoryear{{Ingram}, {Done}, \& {Fragile}}{{Ingram}
  et~al.}{2009}]{2009MNRAS.397L.101I}
{Ingram}, A., {Done}, C.,  \& {Fragile}, P.~C. 2009, \mnras, 397, L101

\bibitem[\protect\citeauthoryear{{Jamil}, {Fender}, \& {Kaiser}}{{Jamil}
  et~al.}{2010}]{2010MNRAS.401..394J}
{Jamil}, O., {Fender}, R.~P.,  \& {Kaiser}, C.~R. 2010, \mnras, 401, 394

\bibitem[\protect\citeauthoryear{{Jonas} \& {MeerKAT Team}}{{Jonas} \& {MeerKAT
  Team}}{2016}]{2016jonas}
{Jonas}, J.,  \& {MeerKAT Team}. 2016, in Proceedings of MeerKAT Science: On
  the Pathway to the SKA. 25-27 May, 1

\bibitem[\protect\citeauthoryear{{Kaaret} et~al.}{{Kaaret}
  et~al.}{2003}]{2003ApJ...582..945K}
{Kaaret}, P., {Corbel}, S., {Tomsick}, J.~A., {Fender}, R., {Miller}, J.~M.,
  {Orosz}, J.~A., {Tzioumis}, A.~K.,  \& {Wijnands}, R. 2003, \apj, 582, 945

\bibitem[\protect\citeauthoryear{{Kalemci} et~al.}{{Kalemci}
  et~al.}{2013}]{2013ApJ...779...95K}
{Kalemci}, E., {Din{\c c}er}, T., {Tomsick}, J.~A., {Buxton}, M.~M., {Bailyn},
  C.~D.,  \& {Chun}, Y.~Y. 2013, \apj, 779, 95

\bibitem[\protect\citeauthoryear{{Kennea} et~al.}{{Kennea}
  et~al.}{2017}]{2017ATel10700....1K}
{Kennea}, J.~A., {Evans}, P.~A., {Beardmore}, A.~P., {Krimm}, H.~A., {Romano},
  P., {Yamaoka}, K., {Serino}, M.,  \& {Negoro}, H. 2017, The Astronomer's
  Telegram, 10700

\bibitem[\protect\citeauthoryear{{King} et~al.}{{King}
  et~al.}{2016}]{2016NatPh..12..772K}
{King}, A.~L., {Miller}, J.~M., {Bietenholz}, M., {G{\"u}ltekin}, K.,
  {Reynolds}, M.~T., {Mioduszewski}, A., {Rupen}, M.,  \& {Bartel}, N. 2016,
  Nature Physics, 12, 772

\bibitem[\protect\citeauthoryear{{Klein-Wolt} \& {van der Klis}}{{Klein-Wolt}
  \& {van der Klis}}{2008}]{2008ApJ...675.1407K}
{Klein-Wolt}, M.,  \& {van der Klis}, M. 2008, \apj, 675, 1407

\bibitem[\protect\citeauthoryear{{Koljonen} et~al.}{{Koljonen}
  et~al.}{2010}]{2010MNRAS.406..307K}
{Koljonen}, K.~I.~I., {Hannikainen}, D.~C., {McCollough}, M.~L., {Pooley},
  G.~G.,  \& {Trushkin}, S.~A. 2010, \mnras, 406, 307

\bibitem[\protect\citeauthoryear{{Koljonen} et~al.}{{Koljonen}
  et~al.}{2018}]{2018A&A...612A..27K}
{Koljonen}, K.~I.~I., {Maccarone}, T., {McCollough}, M.~L., {Gurwell}, M.,
  {Trushkin}, S.~A., {Pooley}, G.~G., {Piano}, G.,  \& {Tavani}, M. 2018, \aap,
  612, A27

\bibitem[\protect\citeauthoryear{{Krimm} et~al.}{{Krimm}
  et~al.}{2013}]{2013ApJS..209...14K}
{Krimm}, H.~A., et~al. 2013, \apjs, 209, 14

\bibitem[\protect\citeauthoryear{{Kylafis} \& {Belloni}}{{Kylafis} \&
  {Belloni}}{2015}]{2015A&A...574A.133K}
{Kylafis}, N.~D.,  \& {Belloni}, T.~M. 2015, \aap, 574, A133

\bibitem[\protect\citeauthoryear{{Kylafis} et~al.}{{Kylafis}
  et~al.}{2012}]{2012A&A...538A...5K}
{Kylafis}, N.~D., {Contopoulos}, I., {Kazanas}, D.,  \& {Christodoulou}, D.~M.
  2012, \aap, 538, A5

\bibitem[\protect\citeauthoryear{{Lepingwell} et~al.}{{Lepingwell}
  et~al.}{2018}]{2018ATel11884....1L}
{Lepingwell}, V.~A., {Bazzano}, A., {Bird}, A.~J., {Chenevez}, J., {Fiocchi},
  M.,  \& {Sguera}, V. 2018, The Astronomer's Telegram, 11884

\bibitem[\protect\citeauthoryear{{Lisakov} et~al.}{{Lisakov}
  et~al.}{2017}]{2017MNRAS.468.4478L}
{Lisakov}, M.~M., {Kovalev}, Y.~Y., {Savolainen}, T., {Hovatta}, T.,  \&
  {Kutkin}, A.~M. 2017, \mnras, 468, 4478

\bibitem[\protect\citeauthoryear{{Liska} et~al.}{{Liska}
  et~al.}{2018}]{2018MNRAS.474L..81L}
{Liska}, M., {Hesp}, C., {Tchekhovskoy}, A., {Ingram}, A., {van der Klis}, M.,
  \& {Markoff}, S. 2018, \mnras, 474, L81

\bibitem[\protect\citeauthoryear{{Maccarone}}{{Maccarone}}{2003}]{2003A&A...409..697M}
{Maccarone}, T.~J. 2003, \aap, 409, 697

\bibitem[\protect\citeauthoryear{{Malzac}}{{Malzac}}{2013}]{2013MNRAS.429L..20M}
{Malzac}, J. 2013, \mnras, 429, L20

\bibitem[\protect\citeauthoryear{{Markoff}, {Falcke}, \& {Fender}}{{Markoff}
  et~al.}{2001}]{2001A&A...372L..25M}
{Markoff}, S., {Falcke}, H.,  \& {Fender}, R. 2001, \aap, 372, L25

\bibitem[\protect\citeauthoryear{{Markoff}, {Nowak}, \& {Wilms}}{{Markoff}
  et~al.}{2005}]{2005ApJ...635.1203M}
{Markoff}, S., {Nowak}, M.~A.,  \& {Wilms}, J. 2005, \apj, 635, 1203

\bibitem[\protect\citeauthoryear{{Markwardt} et~al.}{{Markwardt}
  et~al.}{2017}]{2017GCN.21788....1M}
{Markwardt}, C.~B., {Burrows}, D.~N., {Cummings}, J.~R., {Kennea}, J.~A.,
  {Marshall}, F.~E., {Page}, K.~L., {Palmer}, D.~M.,  \& {Siegel}, M.~H. 2017,
  GRB Coordinates Network, Circular Service, No.~21788, \#1 (2017), 21788

\bibitem[\protect\citeauthoryear{{Marscher} et~al.}{{Marscher}
  et~al.}{2008}]{2008Natur.452..966M}
{Marscher}, A.~P., et~al. 2008, \nat, 452, 966

\bibitem[\protect\citeauthoryear{{Mart{\'\i}-Vidal} et~al.}{{Mart{\'\i}-Vidal}
  et~al.}{2014}]{2014A&A...563A.136M}
{Mart{\'\i}-Vidal}, I., {Vlemmings}, W.~H.~T., {Muller}, S.,  \& {Casey}, S.
  2014, \aap, 563, A136

\bibitem[\protect\citeauthoryear{{McMullin} et~al.}{{McMullin}
  et~al.}{2007}]{2007ASPC..376..127M}
{McMullin}, J.~P., {Waters}, B., {Schiebel}, D., {Young}, W.,  \& {Golap}, K.
  2007, in Astronomical Society of the Pacific Conference Series, Vol. 376,
  Astronomical Data Analysis Software and Systems XVI, ed. R.~A. {Shaw},
  F.~{Hill}, \& D.~J. {Bell}, 127

\bibitem[\protect\citeauthoryear{{Migliori} et~al.}{{Migliori}
  et~al.}{2017}]{2017MNRAS.472..141M}
{Migliori}, G., {Corbel}, S., {Tomsick}, J.~A., {Kaaret}, P., {Fender}, R.~P.,
  {Tzioumis}, A.~K., {Coriat}, M.,  \& {Orosz}, J.~A. 2017, \mnras, 472, 141

\bibitem[\protect\citeauthoryear{{Miller} et~al.}{{Miller}
  et~al.}{2018}]{2018ApJ...860L..28M}
{Miller}, J.~M., et~al. 2018, \apjl, 860, L28

\bibitem[\protect\citeauthoryear{{Miller-Jones}, {Fender}, \&
  {Nakar}}{{Miller-Jones} et~al.}{2006}]{2006MNRAS.367.1432M}
{Miller-Jones}, J.~C.~A., {Fender}, R.~P.,  \& {Nakar}, E. 2006, \mnras, 367,
  1432

\bibitem[\protect\citeauthoryear{{Miller-Jones} et~al.}{{Miller-Jones}
  et~al.}{2011}]{2011MNRAS.415..306M}
{Miller-Jones}, J.~C.~A., {Jonker}, P.~G., {Ratti}, E.~M., {Torres}, M.~A.~P.,
  {Brocksopp}, C., {Yang}, J.,  \& {Morrell}, N.~I. 2011, \mnras, 415, 306

\bibitem[\protect\citeauthoryear{{Miller-Jones} et~al.}{{Miller-Jones}
  et~al.}{2012a}]{2012MNRAS.419L..49M}
{Miller-Jones}, J.~C.~A., et~al. 2012a, \mnras, 419, L49

\bibitem[\protect\citeauthoryear{{Miller-Jones} et~al.}{{Miller-Jones}
  et~al.}{2012b}]{2012MNRAS.421..468M}
{Miller-Jones}, J.~C.~A., et~al. 2012b, \mnras, 421, 468

\bibitem[\protect\citeauthoryear{{Miller-Jones} et~al.}{{Miller-Jones}
  et~al.}{2019}]{2019Natur.569..374M}
{Miller-Jones}, J. C.~A., et~al. 2019, \nat, 569, 374

\bibitem[\protect\citeauthoryear{{Mioduszewski} et~al.}{{Mioduszewski}
  et~al.}{2001}]{2001ApJ...553..766M}
{Mioduszewski}, A.~J., {Rupen}, M.~P., {Hjellming}, R.~M., {Pooley}, G.~G.,  \&
  {Waltman}, E.~B. 2001, \apj, 553, 766

\bibitem[\protect\citeauthoryear{{Mirabel} et~al.}{{Mirabel}
  et~al.}{2011}]{2011A&A...528A.149M}
{Mirabel}, I.~F., {Dijkstra}, M., {Laurent}, P., {Loeb}, A.,  \& {Pritchard},
  J.~R. 2011, \aap, 528, A149

\bibitem[\protect\citeauthoryear{{Mirabel} \& {Rodr{\'{\i}}guez}}{{Mirabel} \&
  {Rodr{\'{\i}}guez}}{1994}]{1994Natur.371...46M}
{Mirabel}, I.~F.,  \& {Rodr{\'{\i}}guez}, L.~F. 1994, \nat, 371, 46

\bibitem[\protect\citeauthoryear{{Mirabel} \& {Rodr{\'{\i}}guez}}{{Mirabel} \&
  {Rodr{\'{\i}}guez}}{1999}]{1999ARA&A..37..409M}
{Mirabel}, I.~F.,  \& {Rodr{\'{\i}}guez}, L.~F. 1999, \araa, 37, 409

\bibitem[\protect\citeauthoryear{{Nakahira} et~al.}{{Nakahira}
  et~al.}{2018}]{2018PASJ...70...95N}
{Nakahira}, S., et~al. 2018, \pasj, 70, 95

\bibitem[\protect\citeauthoryear{{Narayan} \& {Yi}}{{Narayan} \&
  {Yi}}{1995}]{1995ApJ...452..710N}
{Narayan}, R.,  \& {Yi}, I. 1995, \apj, 452, 710

\bibitem[\protect\citeauthoryear{{Negoro} et~al.}{{Negoro}
  et~al.}{2017a}]{2017ATel10699....1N}
{Negoro}, H., et~al. 2017a, The Astronomer's Telegram, 10699

\bibitem[\protect\citeauthoryear{{Negoro} et~al.}{{Negoro}
  et~al.}{2017b}]{2017ATel10708....1N}
{Negoro}, H., et~al. 2017b, The Astronomer's Telegram, 10708

\bibitem[\protect\citeauthoryear{{Negoro} et~al.}{{Negoro}
  et~al.}{2010}]{2010ASPC..434..127N}
{Negoro}, H., et~al. 2010, in Astronomical Society of the Pacific Conference
  Series, Vol. 434, Astronomical Data Analysis Software and Systems XIX, ed.
  Y.~{Mizumoto}, K.-I. {Morita}, \& M.~{Ohishi}, 127

\bibitem[\protect\citeauthoryear{{Negoro} et~al.}{{Negoro}
  et~al.}{2018}]{2018ATel11682....1N}
{Negoro}, H., et~al. 2018, The Astronomer's Telegram, 11682

\bibitem[\protect\citeauthoryear{{Offringa}}{{Offringa}}{2010}]{2010offringa}
{Offringa}, A.~R. 2010, {AOFlagger: RFI Software}

\bibitem[\protect\citeauthoryear{{Parikh} et~al.}{{Parikh}
  et~al.}{2018}]{2018ATel11652....1P}
{Parikh}, A.~S., {Russell}, T.~D., {Wijnands}, R., {Bahramain}, A.,
  {Miller-Jones}, J.~C.~A., {Tetarenko}, A.~J.,  \& {Sivakoff}, G.~R. 2018, The
  Astronomer's Telegram, 11652

\bibitem[\protect\citeauthoryear{{Parikh} et~al.}{{Parikh}
  et~al.}{2019}]{2019ApJ...878L..28P}
{Parikh}, A.~S., {Russell}, T.~D., {Wijnands}, R., {Miller-Jones}, J.~C.~A.,
  {Sivakoff}, G.~R.,  \& {Tetarenko}, A.~J. 2019, \apjl, 878, L28

\bibitem[\protect\citeauthoryear{{Petrucci} et~al.}{{Petrucci}
  et~al.}{2008}]{2008MNRAS.385L..88P}
{Petrucci}, P.-O., {Ferreira}, J., {Henri}, G.,  \& {Pelletier}, G. 2008,
  \mnras, 385, L88

\bibitem[\protect\citeauthoryear{{Rapisarda}, {Ingram}, \& {van der
  Klis}}{{Rapisarda} et~al.}{2014}]{2014MNRAS.440.2882R}
{Rapisarda}, S., {Ingram}, A.,  \& {van der Klis}, M. 2014, \mnras, 440, 2882

\bibitem[\protect\citeauthoryear{{Rees}}{{Rees}}{1966}]{1966Natur.211..468R}
{Rees}, M.~J. 1966, \nat, 211, 468

\bibitem[\protect\citeauthoryear{{Remillard} et~al.}{{Remillard}
  et~al.}{2002}]{2002ApJ...580.1030R}
{Remillard}, R.~A., {Muno}, M.~P., {McClintock}, J.~E.,  \& {Orosz}, J.~A.
  2002, \apj, 580, 1030

\bibitem[\protect\citeauthoryear{{Rodr{\'{\i}}guez} \&
  {Mirabel}}{{Rodr{\'{\i}}guez} \& {Mirabel}}{1999}]{1999ApJ...511..398R}
{Rodr{\'{\i}}guez}, L.~F.,  \& {Mirabel}, I.~F. 1999, \apj, 511, 398

\bibitem[\protect\citeauthoryear{{Romero} et~al.}{{Romero}
  et~al.}{2017}]{2017SSRv..207....5R}
{Romero}, G.~E., {Boettcher}, M., {Markoff}, S.,  \& {Tavecchio}, F. 2017,
  \ssr, 207, 5

\bibitem[\protect\citeauthoryear{{Rupen}, {Hjellming}, \&
  {Mioduszewski}}{{Rupen} et~al.}{1998}]{1998IAUC.6938....2R}
{Rupen}, M.~P., {Hjellming}, R.~M.,  \& {Mioduszewski}, A.~J. 1998, \iaucirc,
  6938

\bibitem[\protect\citeauthoryear{{Rushton} et~al.}{{Rushton}
  et~al.}{2017}]{2017MNRAS.468.2788R}
{Rushton}, A.~P., et~al. 2017, \mnras, 468, 2788

\bibitem[\protect\citeauthoryear{{Russell} et~al.}{{Russell}
  et~al.}{2007}]{2007MNRAS.376.1341R}
{Russell}, D.~M., {Fender}, R.~P., {Gallo}, E.,  \& {Kaiser}, C.~R. 2007,
  \mnras, 376, 1341

\bibitem[\protect\citeauthoryear{{Russell} et~al.}{{Russell}
  et~al.}{2013a}]{2013MNRAS.429..815R}
{Russell}, D.~M., et~al. 2013a, \mnras, 429, 815

\bibitem[\protect\citeauthoryear{{Russell} et~al.}{{Russell}
  et~al.}{2011}]{2011ApJ...739L..19R}
{Russell}, D.~M., {Miller-Jones}, J.~C.~A., {Maccarone}, T.~J., {Yang}, Y.~J.,
  {Fender}, R.~P.,  \& {Lewis}, F. 2011, \apjl, 739, L19

\bibitem[\protect\citeauthoryear{{Russell} et~al.}{{Russell}
  et~al.}{2013b}]{2013ApJ...768L..35R}
{Russell}, D.~M., et~al. 2013b, \apjl, 768, L35

\bibitem[\protect\citeauthoryear{{Russell} et~al.}{{Russell}
  et~al.}{2017}]{2017ATel10711....1R}
{Russell}, T.~D., {Miller-Jones}, J.~C.~A., {Sivakoff}, G.~R., {Tetarenko},
  A.~J.,  \& {Jacpot Xrb Collaboration}. 2017, The Astronomer's Telegram, 10711

\bibitem[\protect\citeauthoryear{{Russell} et~al.}{{Russell}
  et~al.}{2014}]{2014MNRAS.439.1390R}
{Russell}, T.~D., {Soria}, R., {Miller-Jones}, J.~C.~A., {Curran}, P.~A.,
  {Markoff}, S., {Russell}, D.~M.,  \& {Sivakoff}, G.~R. 2014, \mnras, 439,
  1390

\bibitem[\protect\citeauthoryear{{Scaringi} \& {ASTR211 Students}}{{Scaringi}
  \& {ASTR211 Students}}{2017a}]{2017ATel10702....1S}
{Scaringi}, S.,  \& {ASTR211 Students}. 2017a, The Astronomer's Telegram, 10702

\bibitem[\protect\citeauthoryear{{Scaringi} \& {ASTR211 Students}}{{Scaringi}
  \& {ASTR211 Students}}{2017b}]{2017ATel10704....1S}
{Scaringi}, S.,  \& {ASTR211 Students}. 2017b, The Astronomer's Telegram, 10704

\bibitem[\protect\citeauthoryear{{Shibata} \& {Uchida}}{{Shibata} \&
  {Uchida}}{1986}]{1986PASJ...38..631S}
{Shibata}, K.,  \& {Uchida}, Y. 1986, \pasj, 38, 631

\bibitem[\protect\citeauthoryear{{Sikora} et~al.}{{Sikora}
  et~al.}{2005}]{2005ApJ...625...72S}
{Sikora}, M., {Begelman}, M.~C., {Madejski}, G.~M.,  \& {Lasota}, J.-P. 2005,
  \apj, 625, 72

\bibitem[\protect\citeauthoryear{{Silk} \& {Rees}}{{Silk} \&
  {Rees}}{1998}]{1998A&A...331L...1S}
{Silk}, J.,  \& {Rees}, M.~J. 1998, \aap, 331, L1

\bibitem[\protect\citeauthoryear{{Smirnov} \& {Tasse}}{{Smirnov} \&
  {Tasse}}{2015}]{2015tasse}
{Smirnov}, O.~M.,  \& {Tasse}, C. 2015, \mnras, 449, 2668

\bibitem[\protect\citeauthoryear{{Sreehari} et~al.}{{Sreehari}
  et~al.}{2019}]{2019MNRAS.487..928S}
{Sreehari}, H., {Ravishankar}, B.~T., {Iyer}, N., {Agrawal}, V.~K., {Katoch},
  T.~B., {Mandal}, S.,  \& {Nand i}, A. 2019, \mnras, 487, 928

\bibitem[\protect\citeauthoryear{{Stevens} et~al.}{{Stevens}
  et~al.}{2018}]{2018ApJ...865L..15S}
{Stevens}, A.~L., et~al. 2018, \apjl, 865, L15

\bibitem[\protect\citeauthoryear{{Stirling} et~al.}{{Stirling}
  et~al.}{2001}]{2001MNRAS.327.1273S}
{Stirling}, A.~M., {Spencer}, R.~E., {de la Force}, C.~J., {Garrett}, M.~A.,
  {Fender}, R.~P.,  \& {Ogley}, R.~N. 2001, \mnras, 327, 1273

\bibitem[\protect\citeauthoryear{{Tao} et~al.}{{Tao}
  et~al.}{2018}]{2018MNRAS.480.4443T}
{Tao}, L., et~al. 2018, \mnras, 480, 4443

\bibitem[\protect\citeauthoryear{{Tasse} et~al.}{{Tasse}
  et~al.}{2018}]{2018tasse}
{Tasse}, C., et~al. 2018, \aap, 611

\bibitem[\protect\citeauthoryear{{Tetarenko} et~al.}{{Tetarenko}
  et~al.}{2018}]{2018MNRAS.475..448T}
{Tetarenko}, A.~J., {Freeman}, P., {Rosolowsky}, E.~W., {Miller-Jones},
  J.~C.~A.,  \& {Sivakoff}, G.~R. 2018, \mnras, 475, 448

\bibitem[\protect\citeauthoryear{{Tetarenko} et~al.}{{Tetarenko}
  et~al.}{2017a}]{2017ATel10745....1T}
{Tetarenko}, A.~J., {Russell}, T.~D., {Miller-Jones}, J.~C.~A., {Sivakoff},
  G.~R.,  \& {Jacpot Xrb Collaboration}. 2017a, The Astronomer's Telegram,
  10745

\bibitem[\protect\citeauthoryear{{Tetarenko} et~al.}{{Tetarenko}
  et~al.}{2017b}]{2017MNRAS.469.3141T}
{Tetarenko}, A.~J., et~al. 2017b, \mnras, 469, 3141

\bibitem[\protect\citeauthoryear{{Tingay} et~al.}{{Tingay}
  et~al.}{1995}]{1995Natur.374..141T}
{Tingay}, S.~J., et~al. 1995, \nat, 374, 141

\bibitem[\protect\citeauthoryear{{Tomsick} et~al.}{{Tomsick}
  et~al.}{2014}]{2014ApJ...791...70T}
{Tomsick}, J.~A., {Yamaoka}, K., {Corbel}, S., {Kalemci}, E., {Migliari}, S.,
  \& {Kaaret}, P. 2014, \apj, 791, 70

\bibitem[\protect\citeauthoryear{{Vahdat Motlagh}, {Kalemci}, \&
  {Maccarone}}{{Vahdat Motlagh} et~al.}{2019}]{2019MNRAS.485.2744V}
{Vahdat Motlagh}, A., {Kalemci}, E.,  \& {Maccarone}, T.~J. 2019, \mnras, 485,
  2744

\bibitem[\protect\citeauthoryear{{van der Horst} et~al.}{{van der Horst}
  et~al.}{2013}]{2013MNRAS.436.2625V}
{van der Horst}, A.~J., et~al. 2013, \mnras, 436, 2625

\bibitem[\protect\citeauthoryear{{Verner} et~al.}{{Verner}
  et~al.}{1996}]{1996ApJ...465..487V}
{Verner}, D.~A., {Ferland}, G.~J., {Korista}, K.~T.,  \& {Yakovlev}, D.~G.
  1996, \apj, 465, 487

\bibitem[\protect\citeauthoryear{{Wijnands}, {Homan}, \& {van der
  Klis}}{{Wijnands} et~al.}{1999}]{1999ApJ...526L..33W}
{Wijnands}, R., {Homan}, J.,  \& {van der Klis}, M. 1999, \apjl, 526, L33

\bibitem[\protect\citeauthoryear{{Wilms}, {Allen}, \& {McCray}}{{Wilms}
  et~al.}{2000}]{2000ApJ...542..914W}
{Wilms}, J., {Allen}, A.,  \& {McCray}, R. 2000, \apj, 542, 914

\bibitem[\protect\citeauthoryear{{Xu} et~al.}{{Xu}
  et~al.}{2018}]{2018ApJ...852L..34X}
{Xu}, Y., et~al. 2018, \apjl, 852, L34

\bibitem[\protect\citeauthoryear{{Yang} et~al.}{{Yang}
  et~al.}{2010}]{2010MNRAS.409L..64Y}
{Yang}, J., {Brocksopp}, C., {Corbel}, S., {Paragi}, Z., {Tzioumis}, T.,  \&
  {Fender}, R.~P. 2010, \mnras, 409, L64

\end{thebibliography}

\appendix

\section{Radio data}

\begin{center}
\begin{longtable}{cccccc}
\caption {Radio flux densities of \source. Observation MJDs represent the middle of the observation, where errors represent the observation duration. 1-$\sigma$ flux density errors are uncertainties to the fitted source
model. Upper-limits are 3 times the image rms at the source position.} \label{tab:ATCA_data} \\

\hline 
\multicolumn{1}{c}{Start date} & \multicolumn{1}{c}{MJD}  & \multicolumn{1}{c}{Telescope\footnote{}} & \multicolumn{1}{c}{Central frequency} & \multicolumn{1}{c}{Flux density} & \multicolumn{1}{c}{$\alpha$} \\

\multicolumn{1}{c}{(UT)} & \multicolumn{1}{c}{} & \multicolumn{1}{c}{configuration} & \multicolumn{1}{c}{(GHz)} & \multicolumn{1}{c}{(mJy)} & \\ \hline

\endfirsthead

\multicolumn{6}{c}%
{{\tablename\ \thetable{} -- Continued from previous page. Radio flux densities of \source.}} \\

\hline 

\multicolumn{1}{c}{Start date} & \multicolumn{1}{c}{MJD}  & \multicolumn{1}{c}{Telescope\footnote[9]{}} & \multicolumn{1}{c}{Central frequency} & \multicolumn{1}{c}{Flux density} & \multicolumn{1}{c}{$\alpha$} \\

\multicolumn{1}{c}{(UT)} & \multicolumn{1}{c}{} & \multicolumn{1}{c}{configuration} & \multicolumn{1}{c}{(GHz)} & \multicolumn{1}{c}{(mJy)} & \\ \hline

\endhead
\hline
 \multicolumn{6}{c}{{Continued on next page}} \\ \hline
 \multicolumn{6}{c}{{$^9$\footnotesize{https://www.narrabri.atnf.csiro.au/operations/array\_configurations/configurations.html}}}
\endfoot

\hline
\endlastfoot

2017-09-05 & 58001.48$\pm$0.04 & 1.5A & 5.5 & 7.39$\pm$0.03 & 0.09$\pm$0.02 \\
  & &  & 9.0 & 7.74$\pm$0.05 & \\
2017-09-12 & 58008.57$\pm$0.004 & H168 & 17.0 & 171.69$\pm$2.00 & -0.06$\pm$0.15\\
 & &  & 19.0 & 170.52$\pm$2.00 & \\
2017-09-13 & 58009.57$\pm$0.01 & H168 & 5.5 & 192.00$\pm$1.80 &-0.09$\pm$0.01 \\
  & &  & 9.0 & 186.14$\pm$1.00 & \\
  & 58009.541$\pm$0.003 & & 17.0 & 173.72$\pm$1.00 & \\
  & &  & 19.0 & 173.21$\pm$1.00  &\\
2017-09-14 & 58010.563$\pm$0.003 & H168 & 5.5 & 185.30$\pm$1.20 & -0.06$\pm$0.01\\
  & &  & 9.0 & 184.73$\pm$0.22  &\\
  & 58010.56$\pm$0.02 &  & 17.0 & 179.47$\pm$0.25  &\\
  &  & & 19.0 & 175.14$\pm$0.25  &\\
2017-09-15 & 58011.559$\pm$0.003 & H168 & 5.5 & 166.30$\pm$1.10  &-0.02$\pm$0.01\\
  & &  & 9.0 & 181.66$\pm$0.36  &\\
& 58011.56$\pm$0.01 & & 17.0 & 178.47$\pm$0.25  &\\
  &  & & 19.0 & 175.41$\pm$0.30  &\\
2017-09-16 & 58012.55$\pm$0.01 & H168 & 5.5 & 164.00$\pm$1.50  &0.08$\pm$0.01\\
  & &  & 9.0 & 178.40$\pm$1.90  &\\
  & 58012.53$\pm$0.01 & & 17.0 & 184.15$\pm$0.23  &\\
  & &  & 19.0 & 184.05$\pm$0.34  &\\
2017-09-17 & 58013.553$\pm$0.005 & H168 & 5.5 & 135.40$\pm$1.10  &-0.18$\pm$0.01\\
  & &  & 9.0 & 141.77$\pm$0.58  &\\
  & 58013.55$\pm$0.01& & 17.0 & 122.47$\pm$0.24  &\\
  & &  & 19.0 & 118.64$\pm$0.20  &\\
2017-09-21 & 58017.46$\pm$0.09 & H168 & 5.5 & 150.47$\pm$0.08  &-0.45$\pm$0.01\\
  & &  & 9.0 & 121.30$\pm$2.00  &\\
  & 58017.46$\pm$0.10& & 17.0 & 91.81$\pm$0.08  &\\
  & &  & 19.0 & 85.83$\pm$0.07  &\\
2017-09-23 & 58019.52$\pm$0.01 & H168 & 5.5 & 377.20$\pm$1.20 & -0.46$\pm$0.01\\
  &&   & 9.0 & 324.18$\pm$0.34  &\\
  & 58019.52$\pm$0.02 & &17.0 & 240.22$\pm$0.35  &\\
  & &  & 19.0 & 223.18$\pm$0.47  &\\
2017-09-27 & 58023.42$\pm$0.02 & H168 & 5.5 & 127.50$\pm$0.29  &-0.27$\pm$0.01\\
  & &  & 9.0 & 114.28$\pm$0.15  &\\
  & 58023.41$\pm$0.03& & 17.0 & 95.24$\pm$0.22  &\\
  &  & & 19.0 & 90.59$\pm$0.25  &\\
2017-09-30 & 58026.29$\pm$0.01 & H168 & 5.5 & 29.39$\pm$0.23  &-0.20$\pm$0.01\\
  & &  & 9.0 & 26.84$\pm$0.08  &\\
  & 58026.29$\pm$0.02& & 17.0 & 23.00$\pm$0.05  &\\
  & &  & 19.0 & 23.52$\pm$0.06  &\\
2017-10-05 & 58031.40$\pm$0.01 & H168 & 5.5 & 14.73$\pm$0.22 & -0.12$\pm$0.02\\
  & &  & 9.0 & 13.32$\pm$0.08  &\\
  & 58031.40$\pm$0.02& & 17.0 & 12.87$\pm$0.21  &\\
 & &  & 19.0 & 12.28$\pm$0.23  &\\
2017-10-25 & 58051.33$\pm$0.07 & H168 & 5.5 & 75.91$\pm$0.35 & 0.08$\pm$0.01\\
  &  & & 9.0 & 82.61$\pm$0.12  &\\
  & 58051.388$\pm$0.004& & 17.0 & 85.29$\pm$0.19 & \\
  & &  & 19.0 & 87.20$\pm$0.13 & \\
2017-11-02 & 58059.84$\pm$0.03 & 6A & 5.5 & 42.57$\pm$0.04 & -0.30$\pm$0.01\\
  &  & & 9.0 & 39.00$\pm$0.16 & \\
  & 58059.85$\pm$0.03& & 17.0 & 30.01$\pm$0.12 & \\
  & &  & 19.0 & 29.12$\pm$0.11 & \\
2017-11-23 & 58080.24$\pm$0.01 & 1.5C & 5.5 & 10.54$\pm$0.12 & -0.71$\pm$0.05\\
  & &  & 9.0 & 7.44$\pm$0.17 & \\
2017-12-03 & 58090.78$\pm$0.05 & 6C & 5.5 & 0.97$\pm$0.05 & -0.70$\pm$0.11\\
  &  & & 9.0 & 0.74$\pm$0.03 & \\
  & 58090.79$\pm$0.04 & 6C & 17.0 & 0.35$\pm$0.07 & \\
  & &  & 19.0 & 0.36$\pm$0.09 & \\
2017-12-10 & 58097.80$\pm$0.07 & 6C & 5.5 & 0.63$\pm$0.03 & -0.45$\pm$0.06\\
  & &  & 9.0 & 0.51$\pm$0.02 & \\
  & 58097.82$\pm$0.06& & 17.0 & 0.28$\pm$0.02 & \\
  & &  & 19.0 & 0.44$\pm$0.03 & \\
2017-12-16 & 58103.92$\pm$0.07 & 6C & 5.5 & 4.31$\pm$0.22 & -0.64$\pm$0.07\\
  & &  & 9.0 & 3.53$\pm$0.21 & \\
  & 58103.93$\pm$0.07& & 17.0 & 2.24$\pm$0.15 & \\
  & &  & 19.0 & 1.53$\pm$0.20 & \\
2017-12-23 & 58110.98$\pm$0.06 & 6C & 5.5 & 2.20$\pm$0.04 & -1.0$\pm$0.2\\
  & &  & 9.0 & 1.39$\pm$0.07 & \\
  & 58110.98$\pm$0.05& & 17.0 & 0.56$\pm$0.08 & \\
  &  & & 19.0 & 0.48$\pm$0.09 & \\
2017-12-30 & 58117.93$\pm$0.09 & 6C & 5.5 & 1.21$\pm$0.02 & -0.75$\pm$0.04\\
  &  & & 9.0 & 0.91$\pm$0.03 & \\
  & 58117.98$\pm$0.05 && 17.0 & 0.54$\pm$0.03 & \\
  &  & & 19.0 & 0.39$\pm$0.03 & \\
2018-01-05 & 58123.80$\pm$0.18  & 6C& 5.5 & 0.50$\pm$0.02 & -0.60$\pm$0.09\\
  &  & & 9.0 & 0.45$\pm$0.03 & \\
  & 58123.82$\pm$0.18& & 17.0 & 0.21$\pm$0.03 & \\
  &  & & 19.0 & 0.21$\pm$0.04 & \\
2018-01-12 & 58130.93$\pm$0.11 & 6C & 5.5 & 0.18$\pm$0.02 & -0.89$\pm$0.37\\
                           & &  & 9.0 & 0.11$\pm$0.02 & \\
 & 58130.94$\pm$0.11& & 17.0 & $<$0.12 & \\
  & &  & 19.0 & $<$0.17 & \\
2018-01-21 & 58139.73$\pm$0.14 & 750A & 5.5 & 0.32$\pm$0.04 & -0.26$\pm$0.14\\
  & &  & 9.0 & 0.29$\pm$0.03 & \\
  & 58139.73$\pm$0.13& & 17.0 & 0.25$\pm$0.04 & \\
  &  & & 19.0 & 0.22$\pm$0.04 & \\
2018-01-27 & 58146.09$\pm$0.11 & 750A & 5.5 & 0.26$\pm$0.04 & -0.55$\pm$0.40\\
 & &  & 9.0 & 0.20$\pm$0.03 & \\
  & 58146.09$\pm$0.09& & 17.0 & $<$0.15 & \\
  & &  & 19.0 & $<$0.22 & \\
2018-02-02 & 58151.99$\pm$0.09 & 750A & 5.5 & 0.31$\pm$0.02 & -0.42$\pm$0.13\\
  & &  & 9.0 & 0.28$\pm$0.03 & \\
  & 58151.99$\pm$0.07& & 17.0 & 0.19$\pm$0.03 & \\
  & &  & 19.0 & 0.18$\pm$0.04 & \\
2018-02-12 & 58161.80$\pm$0.07 & 750A & 5.5 & 0.27$\pm$0.06 & -1.2$\pm$0.6\\
  & &  & 9.0 & 0.15$\pm$0.03 & \\
  & 58161.80$\pm$0.05& & 17.0 & $<$0.16 & \\
  &  & & 19.0 & $<$0.22 & \\
2018-02-17 & 58166.92$\pm$0.13 & 750B & 5.5 & $<$0.12 & \\
  & &  & 9.0 & $<$0.06 & \\
 & 58166.94$\pm$0.13& & 17.0 & $<$0.11 & \\
  &  & & 19.0 & $<$0.12 & \\
2018-02-22 & 58172.00$\pm$0.11 & 750B & 5.5 & $<$0.1 & \\
  &  & & 9.0 & $<$0.08 & \\
  & 58171.99$\pm$0.09 && 17.0 & $<$0.15 & \\
  &  & & 19.0 & $<$0.18 & \\
2018-02-27 & 58176.67$\pm$0.18 & 750B & 5.5 & $<$0.11 & \\
 &  & & 9.0 & $<$0.09 & \\
2018-03-11 & 58188.48$\pm$0.04& EW352 & 5.5 & $<$0.14 & \\
  &  & & 9.0 & $<$0.12 & \\
2018-03-17 & 58194.47$\pm$0.05& EW352 & 5.5 & $<$0.14 & \\
  & & & 9.0 & $<$0.11 & \\
2018-04-13 & 58221.77$\pm$0.04 & H168 & 5.5 & $<$0.12 & \\
  &  & & 9.0 & $<$0.11 & \\
2018-04-14 & 58222.83$\pm$0.05 & MeerKAT  & 1.3 & $<$0.06 & \\
2018-04-20 & 58228.79$\pm$0.07 & H168 & 5.5 & $<$0.08 & \\
  & &  & 9.0 & $<$0.12 & \\
2018-04-27 & 58235.92$\pm$0.03 & H168 & 5.5 & $<$0.19 & \\
  &  & & 9.0 & $<$0.18 & \\
  & 58235.93$\pm$0.03 & & 17.0 & $<$0.26 & \\
  &  & & 19.0 & $<$0.28 & \\
2018-05-03 & 58241.93$\pm$0.03 & H168 & 5.5 & $<$0.12 & \\
  & &  & 9.0 & $<$0.13 & \\
2018-05-06 & 58244.92$\pm$0.01 & H168 & 5.5 & $<$0.16 & \\
  & &  & 9.0 & $<$0.17 & \\
2018-05-11 & 58249.91$\pm$0.01 & 6D & 5.5 & $<$0.16 & \\
  &  & & 9.0 & $<$0.16 & \\

\end{longtable}
$^9$\footnotesize{https://www.narrabri.atnf.csiro.au/operations/array\_configurations/configurations.html}

\end{center}

\begin{center}
\begin{longtable}{cccccc}
\caption{Radio flux densities of S2. Observation MJDs represent the middle of the observation, where errors represent the observation duration. We also include non-detection close in time to the detections to emphasis the brightenings at late-times. 1-$\sigma$ errors are uncertainties to the fitted source model. Upper-limits are 3 times the image rms at the target position. } \label{tab:ATCA_data_S2} \\

\hline 
\multicolumn{1}{c}{Start date} & \multicolumn{1}{c}{MJD}  & \multicolumn{1}{c}{Telescope\footnote[9]{}} & \multicolumn{1}{c}{Central frequency} & \multicolumn{1}{c}{Flux density} & \multicolumn{1}{c}{$\alpha$} \\

\multicolumn{1}{c}{(UT)} & \multicolumn{1}{c}{} & \multicolumn{1}{c}{configuration} & \multicolumn{1}{c}{(GHz)} & \multicolumn{1}{c}{(mJy)} & \\ \hline

\endfirsthead

\multicolumn{6}{c}%
{{\tablename\ \thetable{} -- Continued from previous page. Radio flux densities of S2.}} \\

\hline 

\multicolumn{1}{c}{Start date} & \multicolumn{1}{c}{MJD}  & \multicolumn{1}{c}{Telescope\footnote[9]{}} & \multicolumn{1}{c}{Central frequency} & \multicolumn{1}{c}{Flux density} & \multicolumn{1}{c}{$\alpha$} \\

\multicolumn{1}{c}{(UT)} & \multicolumn{1}{c}{} & \multicolumn{1}{c}{configuration} & \multicolumn{1}{c}{(GHz)} & \multicolumn{1}{c}{(mJy)} & \\ \hline

\endhead
\hline
 \multicolumn{6}{c}{{Continued on next page}} \\ \hline
 \multicolumn{6}{c}{{$^9$\footnotesize{https://www.narrabri.atnf.csiro.au/operations/array\_configurations/configurations.html}}}
\endfoot

\hline
\endlastfoot

\hline
2017-12-03 & 58090.78$\pm$0.05 & 6C & 5.5 & 2.87$\pm$0.07 & -0.71$\pm$0.02\\
  &  & & 9.0 & 1.98$\pm$0.03 & \\
  & 58090.79$\pm$0.04& & 17.0 & 1.28$\pm$0.06 & \\
  & &  & 19.0 & 1.23$\pm$0.07 & \\
2017-12-10 & 58097.80$\pm$0.07 & 6C & 5.5 & 0.98$\pm$0.05 & -0.82$\pm$0.03\\
  & &  & 9.0 & 0.65$\pm$0.04 & \\
  & 58097.82$\pm$0.06& & 17.0 & 0.37$\pm$0.02 & \\
  & &  & 19.0 & 0.40$\pm$0.04 & \\
2017-12-16 & 58103.92$\pm$0.07 & 6C & 5.5 & 0.39$\pm$0.06 & -0.8$\pm$0.3\\
  & &  & 9.0 & 0.26$\pm$0.06 & \\
2017-12-23 & 58110.98$\pm$0.06 & 6C & 5.5 & 0.45$\pm$0.05 & -0.5$\pm$0.1\\
  &  & & 9.0 & 0.40$\pm$0.08 & \\
  & 58110.98$\pm$0.05& & 17.0 & 0.27$\pm$0.09 & \\
  &  & & 19.0 & 0.22$\pm$0.09 & \\
2017-12-30 & 58117.93$\pm$0.09 & 6C & 5.5 & 0.19$\pm$0.02 & -0.48$\pm$0.08\\
  & &  & 9.0 & 0.16$\pm$0.02 & \\
  & 58117.98$\pm$0.05& & 17.0 & 0.11$\pm$0.03 & \\
  & & & 19.0 & 0.09$\pm$0.03 & \\
2018-01-05 & 58123.80$\pm$0.18 & 6C & 5.5 & 0.17$\pm$0.02 & -0.55$\pm$0.22\\
  &  & & 9.0 & 0.13$\pm$0.03 & \\
  & 58123.82$\pm$0.18& & 17.0 & $<$0.09 & \\
 &  & & 19.0 & $<$0.12 & \\
2018-01-12 & 58130.93$\pm$0.11 & 6C & 5.5 & 0.14$\pm$0.02 & -0.8$\pm$0.2\\
  & &  & 9.0 & 0.10$\pm$0.02 & \\
  & 58130.94$\pm$0.11& & 17.0 & $<$0.12 & \\
  & &  & 19.0 & $<$0.18 & \\
2018-01-21 & 58139.73$\pm$0.14 & 750A & 5.5 & 0.17$\pm$0.04 & -1.0$\pm$0.35\\
  & & & 9.0 & 0.10$\pm$0.03 & \\
  & 58139.73$\pm$0.13& & 17.0 & $<$0.12 & \\
  &  & & 19.0 & $<$0.12 & \\
2018-01-27 & 58146.09$\pm$0.11 & 750A & 5.5 & $<$0.12 & \\
  &  & & 9.0 & $<$0.09 & \\
2018-02-02 & 58151.99$\pm$0.09 & 750A & 5.5 & $<$0.06 & \\
  &  & & 9.0 & $<$0.09 & \\

2018-03-17 & 58194.47$\pm$0.05& EW352 & 5.5 & $<$0.14 & \\
  & & & 9.0 & $<$0.11 & \\
2018-04-13 & 58221.77$\pm$0.04 & H168 & 5.5 & $<$0.12 & \\
  &  & & 9.0 & $<$0.11 & \\
2018-04-14 & 58222.83$\pm$0.05 & MeerKAT  & 1.3 & 0.29$\pm$0.05 & \\
2018-04-20 & 58228.79$\pm$0.07 & H168 & 5.5 & $<$0.08 & \\
  & &  & 9.0 & $<$0.12 & \\
2018-04-27 & 58235.92$\pm$0.03 & H168 & 5.5 & $<$0.19 & \\
  &  & & 9.0 & $<$0.18 & \\
  & 58235.93$\pm$0.03 & & 17.0 & $<$0.26 & \\
  &  & & 19.0 & $<$0.28 & \\
2018-05-03 & 58241.93$\pm$0.03 & H168 & 5.5 & $<$0.12 & \\
  & &  & 9.0 & $<$0.13 & \\

2018-05-03 & 58241.93$\pm$0.03  & H168 & 5.5 & $<$0.12 & \\
  &  & & 9.0 & $<$0.13 & \\
2018-05-06 & 58244.92$\pm$0.01& & 5.5 & $<$0.16 & \\
  & &  & 9.0 & $<$0.17 & \\
2018-05-11 & 58249.91$\pm$0.01 & 6D & 5.5 & $<$0.16 & \\
  &  & & 9.0 & $<$0.16 & \\

2018-05-14 & 58252.79$\pm$0.12 & 6D & 5.5 & 0.13$\pm$0.01 & -0.35$\pm$0.25\\
  & &  & 9.0 & 0.11$\pm$0.02 & \\
2018-05-17 & 58255.37$\pm$0.08 & 6D & 5.5 & 0.20$\pm$0.02 & -0.37$\pm$0.25\\
  & &  & 9.0 & 0.17$\pm$0.04 & \\
2018-05-20 & 58258.45$\pm$0.10 & 6D & 5.5 & $<$0.15 & \\
  & &  & 9.0 & $<$0.15 & \\
  
2018-05-31 & 58269.28$\pm$0.03 & 6D & 5.5 & $<$0.12 \\
& && 9.0 & $<$0.09 \\
2018-06-01 & 58270.30$\pm$0.01 & 6D & 5.5 & $<$0.12 \\
&& & 9.0 & $<$0.095 \\
2018-08-05 & 58335.66$\pm$0.03 & H75 & 5.5 & $<$0.48 \\
&& & 9.0 & $<$0.35 \\

2018-10-02 & 58393.36$\pm$0.09 & 750C & 5.5 & 0.15$\pm$0.02 & -1.0$\pm$0.3\\
  & &  & 9.0 & 0.09$\pm$0.02 & \\

2018-10-14 & 58405.09$\pm$0.03 & 6A & 5.5 & $<$0.15 \\
& && 9.0 & $<$0.09 \\

\end{longtable}
$^9$\footnotesize{https://www.narrabri.atnf.csiro.au/operations/array\_configurations/configurations.html}

\end{center}

\newpage

\begin{center}
\begin{longtable}{cccccc}
\caption{Measured (corrected) positions of S2 and the separation from R.A.=15$^{\rm h}$35$^{\rm m}$19.71$^{\rm s}$', Dec=-57$^{\rm d}$13$^{\rm m}$47.58$^{\rm s}$. Errors are statistical errors on the fitted position.} \label{tab:S2_positions} \\

\hline 
\multicolumn{1}{c}{Start date} & \multicolumn{1}{c}{MJD} & \multicolumn{1}{c}{Right Ascension} & \multicolumn{1}{c}{Declination} &  \multicolumn{2}{c}{Separation}\\

\multicolumn{1}{c}{(UT)} & \multicolumn{1}{c}{} & \multicolumn{1}{c}{} & \multicolumn{1}{c}{} & \multicolumn{1}{c}{R.A. (\arcsecond)} & \multicolumn{1}{c}{Dec. (\arcsecond)}  \\

\endfirsthead

\multicolumn{6}{c}%
{{\tablename\ \thetable{} -- Continued from previous page. Measured positions of S2.}} \\

\hline 

\multicolumn{1}{c}{Start date} & \multicolumn{1}{c}{MJD} & \multicolumn{1}{c}{Right Ascension} & \multicolumn{1}{c}{Declination} &  \multicolumn{2}{c}{Separation}\\

\multicolumn{1}{c}{(UT)} & \multicolumn{1}{c}{} & \multicolumn{1}{c}{} & \multicolumn{1}{c}{} & \multicolumn{1}{c}{R.A. (\arcsecond)} & \multicolumn{1}{c}{Dec. (\arcsecond)}  \\
\hline

\endhead
\hline
 \multicolumn{6}{c}{{Continued on next page}} \\ \hline
\endfoot

\hline
\endlastfoot

\hline

2017-12-03 & 58090.78$\pm$0.05 & 15$^{\rm h}$35$^{\rm m}$20.12$^{\rm s}\pm0.18$\arcsecond & -57$^{\circ}$13$^{\prime}$49.92$^{\prime\prime}\pm 0.09$\arcsecond & 3.33$\pm$0.18 & -2.34$\pm$0.09 \\
2017-12-10 & 58097.80$\pm$0.07 & 15$^{\rm h}$35$^{\rm m}$20.15$^{\rm s}\pm0.13$\arcsecond & -57$^{\circ}$13$^{\prime}$50.03$^{\prime\prime}\pm0.06$\arcsecond & 3.61$\pm$0.13 & -2.45$\pm$0.06  \\
2017-12-16 & 58103.92$\pm$0.07 & 15$^{\rm h}$35$^{\rm m}$20.11$^{\rm s}\pm0.10$\arcsecond & -57$^{\circ}$13$^{\prime}$49.64$^{\prime\prime}\pm0.14$\arcsecond & 3.28$\pm$0.10 & -2.06$\pm$0.14 \\
2017-12-23 & 58110.98$\pm$0.06  & 15$^{\rm h}$35$^{\rm m}$20.18$^{\rm s}\pm0.07$\arcsecond & -57$^{\circ}$13$^{\prime}$49.96$^{\prime\prime}\pm0.20$\arcsecond & 3.84$\pm$0.07 & -2.38$\pm$0.20  \\
2018-12-30 & 58117.93$\pm$0.09 & 15$^{\rm h}$35$^{\rm m}$20.21$^{\rm s}\pm0.13$\arcsecond  & -57$^{\circ}$13$^{\prime}$50.25$^{\prime\prime}\pm0.29$\arcsecond & 4.08$\pm$0.13 & -2.67$\pm$0.29 \\
2018-01-05 & 58123.80$\pm$0.18 & 15$^{\rm h}$35$^{\rm m}$20.24$^{\rm s}\pm0.12$\arcsecond & -57$^{\circ}$13$^{\prime}$50.67$^{\prime\prime}\pm0.11$\arcsecond & 4.33$\pm$0.12 & -3.09$\pm$0.11  \\ 
2018-01-12 & 58130.93$\pm$0.11 & 15$^{\rm h}$35$^{\rm m}$20.30$^{\rm s}\pm0.13$\arcsecond & -57$^{\circ}$13$^{\prime}$50.88$^{\prime\prime}\pm0.26$\arcsecond  & 4.78$\pm$0.13 & -3.30$\pm$0.26 \\
2018-01-21 & 58139.73$\pm$0.14 & 15$^{\rm h}$35$^{\rm m}$20.35$^{\rm s}\pm0.23$\arcsecond & -57$^{\circ}$13$^{\prime}$51.45$^{\prime\prime}\pm0.23$\arcsecond & 5.19$\pm$0.23 & -3.87$\pm$0.23\\
2018-04-14 & 58222.83$\pm$0.05 & 15$^{\rm h}$35$^{\rm m}$20.71$^{\rm s}\pm0.95$\arcsecond & -57$^{\circ}$13$^{\prime}$52.41$^{\prime\prime}\pm0.40$\arcsecond & 8.12$\pm$0.95 & -4.83$\pm$0.40\\
2018-05-14 & 58252.79$\pm$0.12 & 15$^{\rm h}$35$^{\rm m}$20.88$^{\rm s}\pm0.12$\arcsecond & -57$^{\circ}$13$^{\prime}$54.16$^{\prime\prime}\pm0.08$\arcsecond & 9.52$\pm$0.12 & -6.58$\pm$0.08 \\
2018-05-17 & 58255.37$\pm$0.12 & 15$^{\rm h}$35$^{\rm m}$20.86$^{\rm s}\pm0.20$\arcsecond & -57$^{\circ}$13$^{\prime}$54.00$^{\prime\prime}\pm0.11$\arcsecond & 9.33$\pm$0.20 & -6.42$\pm$0.11 \\
2018-10-02 & 58393.36$\pm$0.09 & 15$^{\rm h}$35$^{\rm m}$21.40$^{\rm s}\pm0.18$\arcsecond & -57$^{\circ}$13$^{\prime}$56.78$^{\prime\prime}\pm0.16$\arcsecond & 13.72$\pm$0.18 & -9.20$\pm$0.16 \\

\end{longtable}
\end{center}

\newpage

\begin{figure}
\centering
\includegraphics[width=0.8\columnwidth]{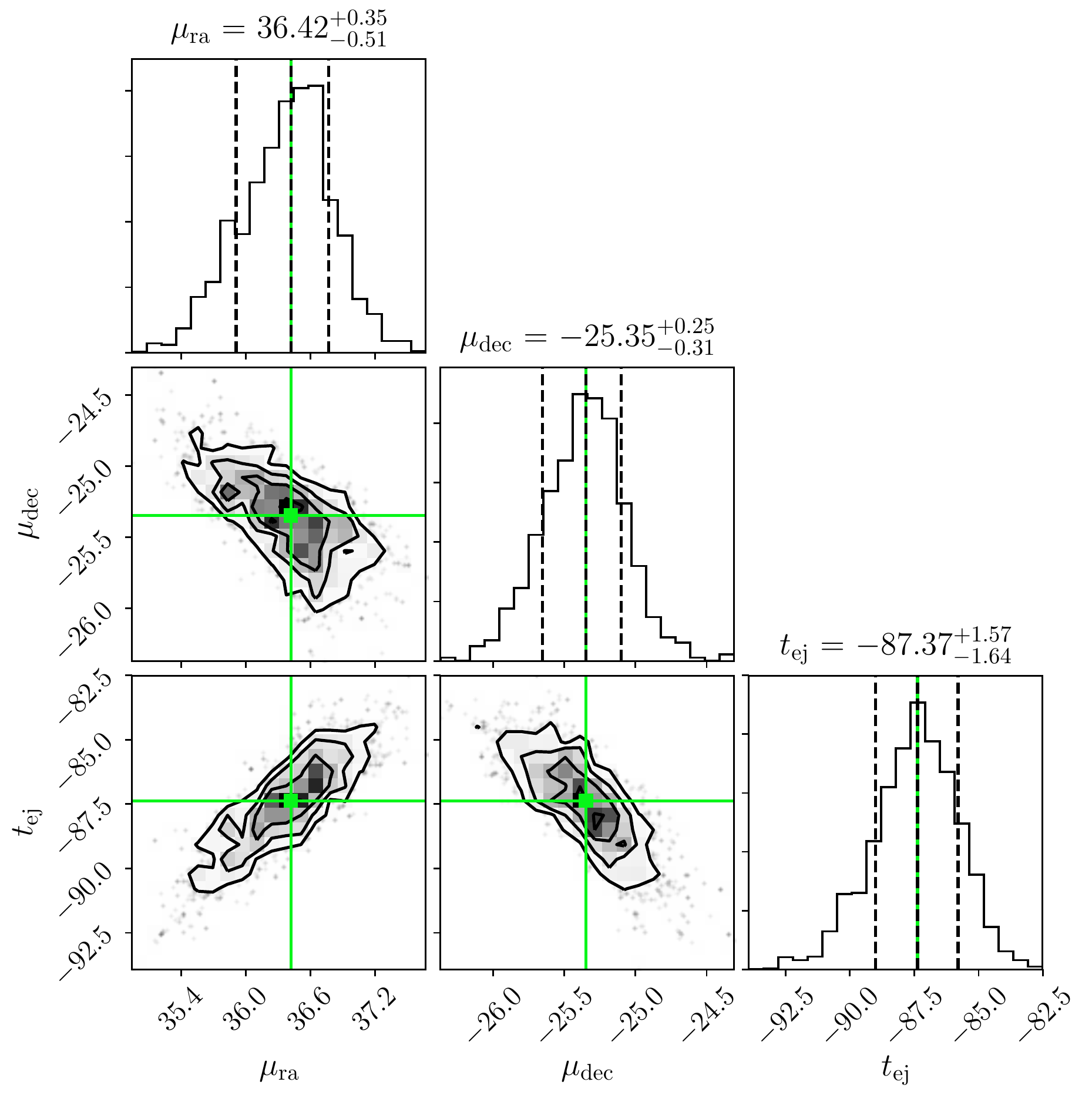}
\caption{Correlation plots of parameters for the simple bulk motion model. Here we show the proper motion in R.A. (${\mu}_{\rm ra}$) and Declination (${\mu}_{\rm dec}$), as well as the best-fit date of the ejection ($t_{\rm ej}$), normalised to the MJD of the first S2 detection (MJD~58090.78). The histograms represent the one dimensional posterior distributions of the parameters, and the green lines/squares indicate the best fit value of the parameters.}
\label{fig:pc1}
\end{figure}

\begin{figure}
\centering
\includegraphics[width=0.8\columnwidth]{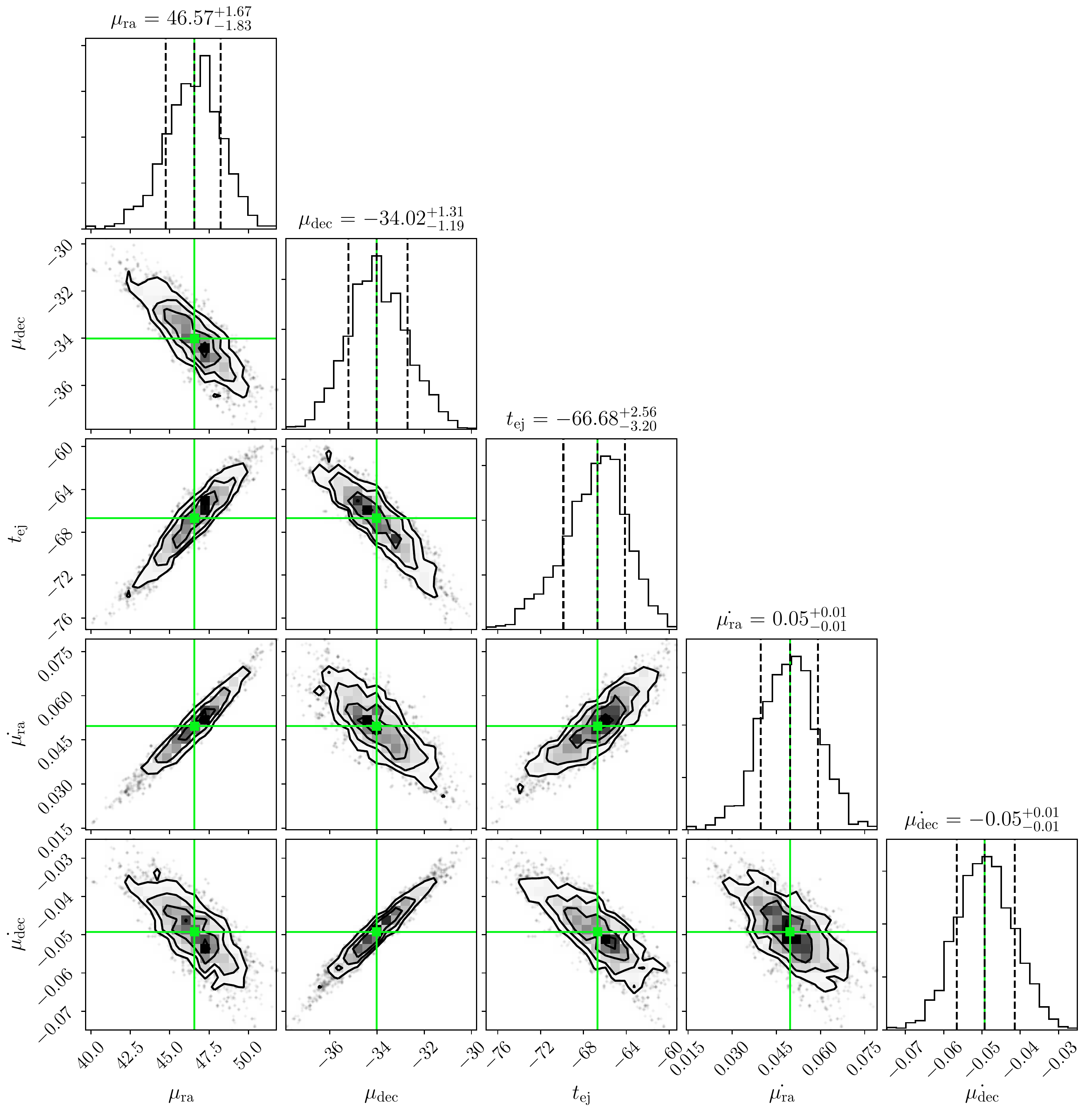}
\caption{Correlation plots of parameters for the deceleration model. Here we show the average proper motion in R.A. (${\mu}_{\rm ra}$) and Declination (${\mu}_{\rm dec}$), the best-fit date of the ejection ($t_{\rm ej}$) normalised to the MJD of the first S2 detection, MJD~58090.78, and the average acceleration in both R.A. and Dec ($\dot{\mu}_{\rm ra}$ and $\dot{\mu}_{\rm dec}$, respectively). The histograms represent the one dimensional posterior distributions of the parameters, and the green lines/squares indicate the best fit value of the parameters.}
\label{fig:pc2}
\end{figure}

\begin{figure}
\centering
\includegraphics[width=0.8\columnwidth]{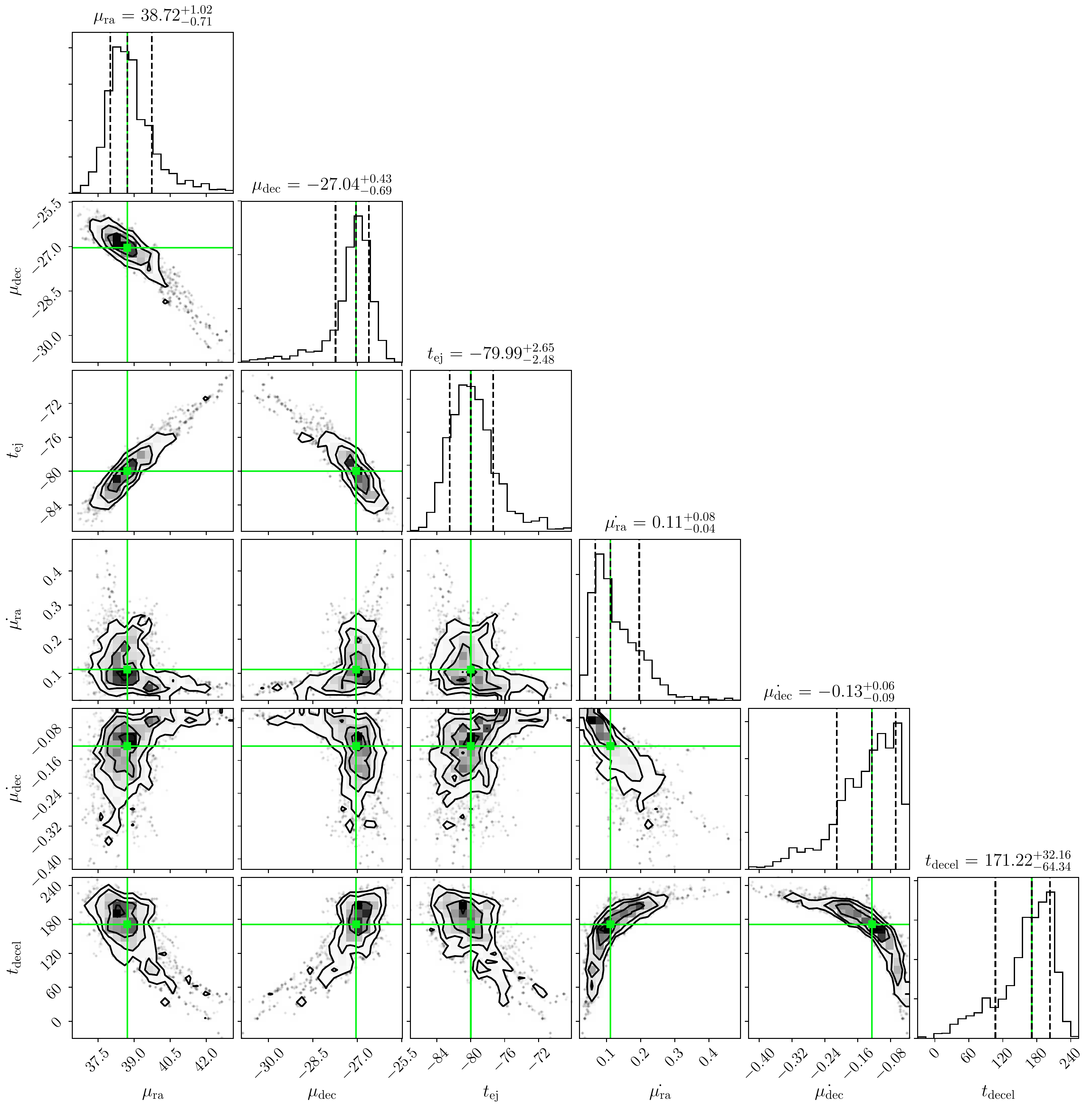}
\caption{Correlation plots of parameters for the combined bulk plus deceleration model. Here we show the average proper motion in R.A. (${\mu}_{\rm ra}$) and Declination (${\mu}_{\rm dec}$), the best-fit date of the ejection ($t_{\rm ej}$) normalised to the MJD of the first S2 detection, MJD~58090.78, the average acceleration in both R.A. and Dec ($\dot{\mu}_{\rm ra}$ and $\dot{\mu}_{\rm dec}$, respectively), and the deceleration start date $t_{\rm decel}$. The histograms represent the one dimensional posterior distributions of the parameters, and the green lines/squares indicate the best fit value of the parameters.}
\label{fig:pc3}
\end{figure}

\end{document}